\documentclass[journal]{IEEEtran}
\usepackage[inline]{trackchanges}

\usepackage[hyphens]{url} 
\usepackage{graphicx} 
\usepackage[square,numbers]{natbib}  
\usepackage{caption} 
\usepackage[ruled,vlined,linesnumbered,commentsnumbered,longend]{algorithm2e}
\usepackage{algorithmic}
\usepackage{amsmath}
\usepackage{longtable}
\usepackage{amsthm}
\usepackage{amssymb}
\usepackage{amsfonts}
\usepackage{bm}
\usepackage{array}
\usepackage{multirow}
\usepackage[english]{babel}
\usepackage{adjustbox}
\usepackage{subfigure}
\usepackage{booktabs}
\usepackage{enumitem}
\usepackage{newfloat}
\usepackage{listings}
\usepackage{url}
\usepackage{verbatim}
\usepackage{xcolor}

\usepackage{epstopdf}
\usepackage{mathrsfs}
\usepackage{url}
\makeatletter
\g@addto@macro{\UrlBreaks}{\UrlOrds}
\makeatother
\usepackage[flushleft]{threeparttable} %
\newtheorem{definition}{Definition}

\newtheorem{problem}{Problem}

\definecolor{cvprblue}{rgb}{0.21,0.49,0.74}
\usepackage[pagebackref,breaklinks,colorlinks,allcolors=cvprblue]{hyperref}

\begin{document}

\title{RouteKG: A knowledge graph-based framework for route prediction on road networks}

\author{Yihong~Tang,
        Zhan~Zhao*,
        Weipeng~Deng, Shuyu~Lei, Yuebing~Liang, Zhenliang~Ma
\thanks{Y. Tang is with McGill University, Montreal, Canada; Z. Zhao, W. Deng and S. Lei are with The University of Hong Kong, Hong Kong SAR, China; Y. Liang is with Tsinghua University, Beijing, China; Z. Ma is with KTH Royal Institute of Technology, Stockholm, Sweden.}
\thanks{* Corresponding author. (E-mail: zhanzhao@hku.hk)}
}

\markboth{IEEE Transactions on Intelligent Transportation Systems, 2025}{}

\maketitle

\begin{abstract}
Short-term route prediction on road networks allows us to anticipate the future trajectories of road users, enabling various applications ranging from dynamic traffic control to personalized navigation. Despite recent advances in this area, existing methods focus primarily on learning sequential transition patterns, neglecting the inherent spatial relations in road networks that can affect human routing decisions. To fill this gap, this paper introduces RouteKG, a novel Knowledge Graph-based framework for route prediction. Specifically, we construct a Knowledge Graph on the road network to encode spatial relations, especially moving directions that are crucial for human navigation. Moreover, an n-ary tree-based algorithm is introduced to efficiently generate top-K routes in batch mode, enhancing computational efficiency. To further optimize prediction performance, a rank refinement module is incorporated to fine-tune candidate route rankings. The model performance is evaluated using two real-world vehicle trajectory datasets from two Chinese cities under various practical scenarios. The results demonstrate a significant improvement in accuracy over the baseline methods. We further validate the proposed method by utilizing the pre-trained model as a simulator for real-time traffic flow estimation at the link level. RouteKG has great potential to transform vehicle navigation, traffic management, and a variety of intelligent transportation tasks, playing a crucial role in advancing the core foundation of intelligent and connected urban systems. The source codes of RouteKG are available at \url{https://github.com/YihongT/RouteKG}.
\end{abstract}



\section{Introduction} \label{sec:intro}

In intelligent transportation and urban systems, with the increasing prevalence of mobile sensors (e.g., GPS devices) and vehicular communication technologies, the ability to predict road users' future routes is not merely a convenience but a necessity to support a range of applications such as vehicle navigation \cite{ziebart_navigate_2008}, traffic management \cite{li_trajectory_2020}, accident prediction \cite{gao2024uncertainty} and location-based recommendation \cite{kong_time-location-relationship_2017,tang2022hgarn}. There are generally two types of route prediction tasks. On the one hand, for transport planning applications, it is often required to predict the complete route (as a sequence of road links) from the origin to the destination. This is typically referred to as \textit{route choice modeling} in the literature \cite{prato_route_2009}, where information about the destination (or goal) must be given. On the other hand, for real-time ITS applications, goal information may not be available, and it is usually adequate to predict the near-future route trajectory of a moving agent based on the observed trajectory so far. This study focuses on the latter, which we call the \textit{short-term route prediction} problem.

Numerous methods have been proposed in the literature to address this problem. Recent works typically use Recurrent Neural Networks (RNNs) \cite{rumelhart1986learning}, especially Long Short-Term Memory (LSTM) \cite{hochreiter1997long} and Gated Recurrent Unit (GRU) \cite{cho2014learning}, to capture sequential dependencies in trajectory data \cite{alahi2016social,mo2023predicting}. Most existing models focus primarily on learning sequential patterns for route prediction, often overlooking the inherent spatial structure of road networks that can affect human routing decisions. To address this issue, some studies have started to use Graph Neural Networks (GNNs) \cite{kipf2016semi} to encode road networks for improved prediction of vehicle trajectories \cite{liang2021nettraj} and traffic conditions \cite{zhao2019t,li2017diffusion}. However, these methods still treat road networks merely as generic graphs, oversimplifying their structure and disregarding crucial spatial factors.

As a type of spatial network, road networks consist of a set of spatial entities ({\em e.g.}, intersections, links, etc.) organized in a way to facilitate traffic flows in a mostly 2-dimensional space. The relationships between these entities can be described by a set of spatial factors such as direction, distance, and connectivity. For example, one of the important spatial factors to consider in routing problems is the direction of travel ({\em i.e.}, goal direction). It has been widely recognized in the navigation and cognitive psychology literature that humans utilize directional cues to navigate their environment \cite{etienne2004path,chrastil2015active}. Existing short-term route prediction models, however, often neglect the spatial structural information of road networks and human-intended moving directions, thus possibly leading to sub-optimal model performance. These limitations highlight the need for a more spatially explicit model to incorporate such spatial relationships that influence human routing behavior.

There are other challenges for short-term route prediction on road networks. Firstly, most existing methods focus on generating a single predicted route \cite{rathore2019scalable,yan2022precln}. However, due to the inherent uncertainties, providing multiple route predictions can have more practical implications. For instance, traffic managers can optimize real-time traffic flow by considering multiple potential routes of moving vehicles, and road users can benefit from having a wider variety of routing options. Secondly, as road networks grow in size, scalability becomes a challenge for GNN-based methods \cite{hamilton2017inductive}, as they require substantial computational and memory resources. Lastly, model prediction performance is heavily dependent on the availability of goal information. Generally, better performance can be achieved by incorporating more goal information into the model. However, the availability of goal information varies, including (1) no information available, (2) only goal direction known, and (3) complete goal information known. These varying degrees of goal information availability can affect route prediction to different extents, but a comprehensive evaluation across all these scenarios is missing.

With the aforementioned challenges, we propose a novel model, \textit{RouteKG}, which leverages Knowledge Graphs (KGs) \cite{wang2014knowledge} to encode road networks for short-term route prediction. Unlike existing models that rely on sequence-to-sequence (seq2seq) structures \cite{sutskever2014sequence}, 
RouteKG interprets route prediction as a Knowledge Graph Completion (KGC) task \cite{chen2020knowledge}. 
Specifically, we propose a \textit{Knowledge Graph Module} that can predict the future links (tail entities) a user might traverse based on current links (head entities) and moving directions (relations) without solely relying on seq2seq structures. The module explicitly incorporates the goal moving direction (estimated or actual) into the future route prediction process, better aligning with the intrinsic nature of human navigation. In addition, we employ a \textit{Route Generation Module} to efficiently generate top-$K$ route candidates, and a \textit{Rank Refinement Module} that can model the dependencies between different links within each predicted route to rerank the route candidates for consistency, resulting in the final top-$K$ predictions. Our proposed KG-based approach can effectively model the spatial relations, thus outperforming existing baselines by a large margin, and it could benefit a range of transportation or routing tasks.
The contributions of this study are summarized as follows:
\begin{itemize}[leftmargin=*,noitemsep]
    \item We introduce RouteKG, a novel KG-based modeling framework for short-term route prediction. In this approach, we adapt the KG to represent road networks and reformulate the route prediction problem as a KGC task. Then, the learned road network and route representations can enhance model prediction and interpretability.
    \item We propose an $n$-ary tree-based route generation algorithm that enables efficient batch generation of future routes based on predicted probabilities derived from the KG. Additionally, we employ a rank refinement module that effectively prioritizes routes for their consistency by modeling dependencies between road links, resulting in more accurate, trustworthy, and reliable top-$K$ route predictions.
    \item Through extensive experiments on two real-world vehicle trajectory datasets from Chengdu and Shanghai, the results validate the superior prediction performance of RouteKG over state-of-the-art baseline models across various scenarios of goal information availability, with low response latency. 
    Furthermore, a case study using the trained RouteKG as a simulator to estimate real-time traffic flows at the link level demonstrates our method's effectiveness in diverse application scenarios.
\end{itemize}

\section{Literature Review} \label{sec:relatedwork}

\subsection{Trajectory Prediction}

\subsubsection{Motion Prediction}

Motion prediction, which anticipates an agent's future trajectory from past movements, is central to autonomous driving systems \cite{yurtsever2020survey, lefevre2014survey}. Its importance has amplified with advancements in autonomous driving and robot navigation, improving safety and efficiency by mitigating collision risks and boosting performance \cite{rudenko2020human}. However, the dynamic and uncertain nature of agents' movements presents unique challenges \cite{paravarzar2020motion}.
Motion prediction methods can generally be divided into two broad categories: classic and deep learning-based, each with unique advantages and limitations.

Classic methods utilize mathematical models grounded in physics and geometry to focus on the deterministic aspects of an agent's motion, offering simplicity, interpretability, and efficiency \cite{helbing1995social}. Yet, these methods struggle to capture the stochastic behavior of agents in complex environments \cite{huang2022survey}.

On the other hand, deep learning-based motion prediction methods use neural networks to model the complexities of agent behavior \cite{alahi2016social}. These methods aim to learn the intricate, often non-linear, relationships between different influencing factors from large-scale data. Approaches such as RNNs and Generative Adversarial Networks (GANs) are commonly used \cite{gupta2018social,sadeghian2019sophie,gu2021densetnt}. Recent efforts employ diffusion process \cite{ho2020denoising} simulate the process of human motion variation from indeterminate to determinate \cite{gu2022stochastic}. The advantage of deep learning methods is their ability to capture the underlying patterns and subtleties that traditional mathematical models might miss. However, they require extensive computational resources and large amounts of training data, and often lack the interpretability of classic methods \cite{rudenko2020human}.

\subsubsection{Route Prediction}

Route prediction, distinct from motion prediction, forecasts the future trajectories of agents that typically operate within road network constraints, necessitating different problem formulations and solutions. Similar to motion prediction, models designed for short-term route prediction can also be broadly classified into traditional approaches and deep learning-based methods.
Traditional methods utilize shortest path-based methods such as Dijkstra's algorithm \cite{dijkstra1959note}, Bellman-Ford, and A* \cite{hart1968formal} for route prediction tasks. However, these dynamic programming-based methods require destination information to generate potential routes for trajectory prediction. As the destination information is often unavailable for short-term route prediction, other works have employed Kalman Filters \cite{abbas2020adaptive} or Hidden Markov Models (HMMs) \cite{simmons2006learning,ye2016vehicle} to predict users' destinations and routes. Nevertheless, these methods struggle to model long-term temporal dependencies due to relatively simple model structures.

In comparison, deep learning-based methods have outperformed traditional methods in prediction tasks, exhibiting superior ability in modeling spatial-temporal dependencies. The RNN-based encoder-decoder trajectory representation learning framework \cite{fu2020trembr} can adapt to tasks such as trajectory similarity measurement, travel time prediction, and destination prediction. Other studies have utilized Graph Convolutional Networks (GCN) and attention mechanisms to refine trajectory representation for prediction purposes \cite{shao2021trajforesee}. Some studies have proposed models for tasks ranging from predicting the next link using historical trajectories \cite{liu2022modeling} to enhancing route prediction through pre-training and contrastive learning \cite{yan2022precln}. Furthermore, some models are designed for road network-constrained trajectory recovery, capable of recovering fine-grained points from low-sampling records \cite{ren2021mtrajrec,chen2022rntrajrec}.

Despite significant progress in short-term route prediction on road networks, many existing methods view it as a sequence-to-sequence task, leveraging sequential models like RNNs or Transformers for prediction. These methods often overlook the crucial role of spatial relations within the road network, an essential aspect of routing tasks.

\subsection{Knowledge Graph}

\subsubsection{Knowledge Graph Completion}

The rapidly expanding interest in KGs has fueled the advancement in tasks like recommender systems, question answering, and semantic search, given their ability to provide structured and machine-interpretable knowledge about real-world entities and their relations \cite{noy2019industry,sheth2019knowledge,paulheim2017knowledge}. Despite their immense potential, a critical problem is the inherent incompleteness of information, making KGC an important and burgeoning research area. KGC refers to inferring missing or incomplete information in a KG by predicting new relationships between entities based on existing information \cite{chen2020knowledge}.

Earlier studies on KGC typically employed statistical relational learning (SRL) methods, such as Markov Logic Networks (MLN) \cite{richardson2006markov} and Probabilistic Soft Logic (PSL) \cite{bach2017hinge}. These methods demonstrate effectiveness in capturing complex dependencies but need to improve scalability due to the need to specify all possible rules manually. More scalable machine learning approaches, especially those involving embeddings, have been proposed to overcome these limitations in recent years. TransE is a seminal model in this line, which models relations as translations in the entity embedding space \cite{bordes2013translating}. Follow-up models such as TransH \cite{wang2014knowledge}, TransR \cite{lin2015learning}, and TransD \cite{ji2015knowledge} were subsequently proposed to handle complex relational data by introducing hyperplanes, relation-specific spaces, or dynamic mapping matrices respectively. Meanwhile, tensor factorization-based models like RESCAL \cite{nickel2011three}, DistMult \cite{yang2014embedding}, and ComplEx \cite{trouillon2016complex} have been developed, aiming to capture the complex correlations between entities and relations. These models generally perform well but can be computationally intensive. More recently, models based on GNNs have shown promising results for KGC. Models such as R-GCN \cite{schlichtkrull2018modeling} and CompGCN \cite{vashishth2019composition} have achieved competitive results by modeling KGs as multi-relational graphs and learning from both the graph structure and node attributes.

To summarize, KGC is a process that leverages machine learning to infer and predict missing knowledge automatically. It utilizes the rich structure of KGs, employing effective entity and relation representations for improved prediction.

\subsubsection{Mobility Knowledge Graph}

KGs have been increasingly utilized to address complex urban mobility problems. Mobility KGs have witnessed considerable growth and advancements in recent years, particularly with the integration of multi-source transportation data, creating KGs derived from GPS trajectory data, and utilizing structured knowledge bases to augment urban mobility data analysis.
\cite{tan2021research} devised a KG for urban traffic systems to uncover the implicit relationships amongst traffic entities and thereby unearth valuable traffic knowledge. Similarly, \cite{zhuang2017understanding} constructed an urban movement KG using GPS trajectory data and affirmed the practicality of their model by predicting the level of user attention directed towards various city locations. \cite{zhao2020urban} put forth a generalized framework for multi-source spatiotemporal data analysis, underpinned by KG embedding, intending to discern the network structure and semantic relationships embedded within multi-source spatiotemporal data. Several studies have focused on building KGs grounded on geographical information and human mobilities for various applications, such as predicting subsequent locations (i.e., Point of Interest recommendation) \cite{liu2021improving,rao2022graph,wang2021spatio}, modeling event streams \cite{wang2020incremental}, learning user similarity \cite{zhang_user_2023}, forecasting destinations \cite{li2022potential,chi2022knowledge}, and performing epidemic contact tracing \cite{chen2022knowledge}.

Despite their methodological divergence, these approaches rely on different data sources to construct mobility KGs, often resulting in superior outcomes but potentially sacrificing some generalizability. Notably, current work has yet to address the design of KGs for route prediction or road network representation learning while retaining generalizability.

\begin{figure*}[ht]
    \centering
    \includegraphics[width=.6\textwidth]{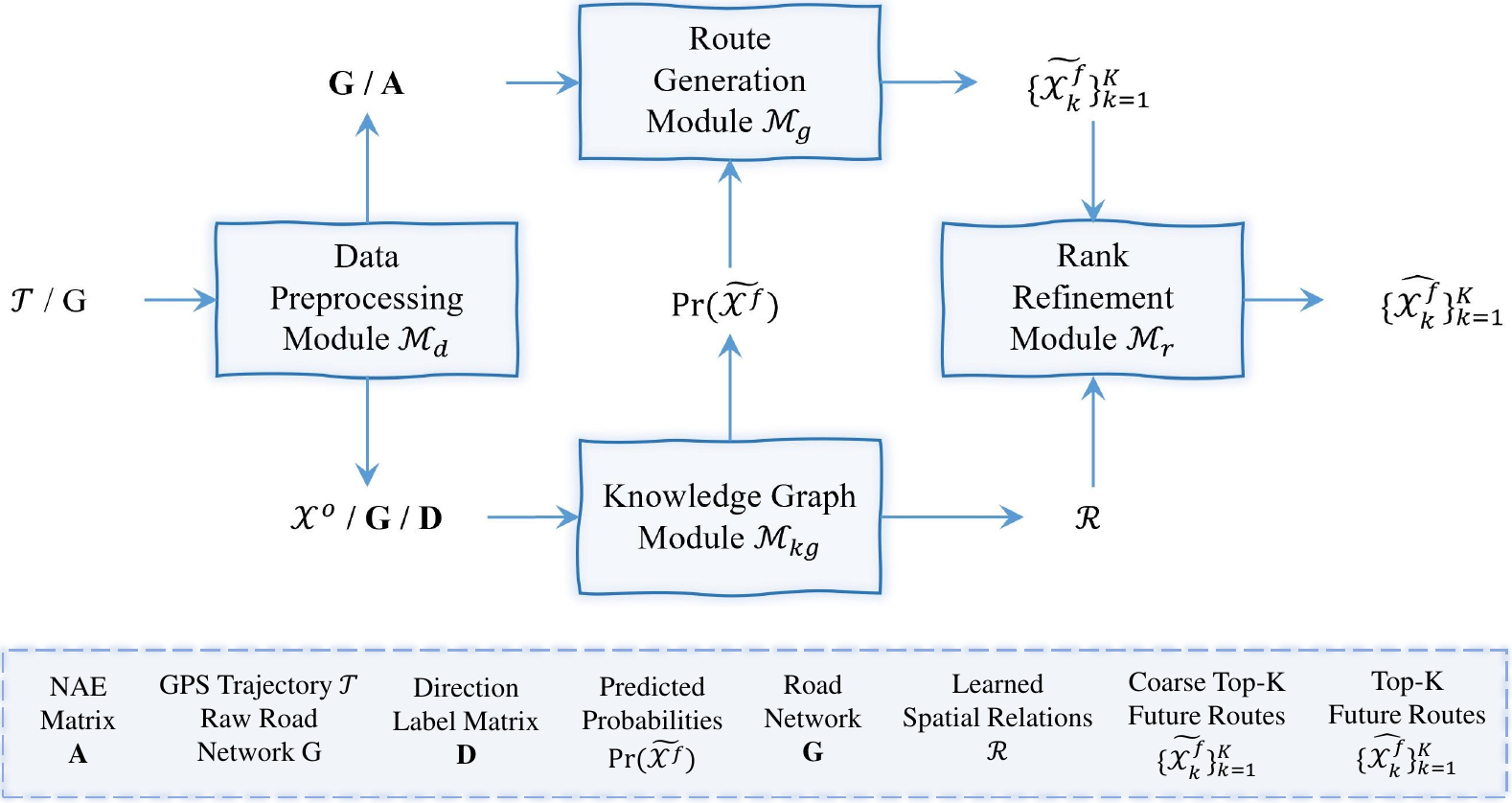}
    \caption{The workflow of RouteKG.}
    \label{fig:flow}
    \vspace{-3mm}
\end{figure*}

\section{Preliminaries} \label{sec:preliminary}

In this section, we introduce definitions and the problem formulation in Section \ref{subsec:pre:problem}. 

\subsection{Problem Formulation} \label{subsec:pre:problem}

\begin{definition}[Road Network $\mathbf{G}$]
The road network can be modeled as a Multi-Directed Graph (MultiDiGraph) $\mathbf{G}=(\mathbf{V}, \mathbf{E})$, where $\mathbf{V}$ is a set of vertices (or nodes) representing unique intersections or endpoints in the road network, and $\mathbf{E}$ is a set of directed edges, each representing a link. Each vertex $v \in \mathbf{V}$ is associated with a geographical coordinate $(lat_v, lon_v)$. Each edge $e \in \mathbf{E}$ carries certain attributes, such as length, road type, etc. Multiple edges may connect the same pair of vertices, accounting for multiple links connecting the same intersections ({\em e.g.}, parallel roads). An edge $e^k$ is denoted as $e^k=(v^s_k, v^e_k, m)$, where $m$ distinguishes edges connecting the same pair of nodes.
\end{definition}

\begin{definition}[Route $x$]
We denote the full set of map-matched routes as $\mathcal{X} = \{x_i\}_{i=1}^{|\mathcal{X}|}$. Each route $x \in \mathcal{X}$ is derived from raw GPS trajectories via map-matching \cite{Yang2018FastMM}. A map-matched route $x$ of length $m$ is thus a sequence of road links, $x = \left\{e^1, e^2, ..., e^m\right\},\ x \in \mathcal{X}$. For every consecutive pair of links $(e^i, e^{i+1})$, there exists a node $v$ in the road graph $\mathbf{G}$ such that $v$ connects the two edges.
The $i$-th route can be partitioned into an observed route $x^{o}_{i}=\{e^j_i\}^{\Gamma}_{j=1}$ of length $\Gamma$ and a future route $x^{f}_{i}=\{e^j_i\}_{j=\Gamma+1}^{\Gamma+\Gamma^\prime}$ of length $\Gamma^\prime$, where both $\Gamma$ and $\Gamma^\prime$ are fixed. The full sets of observed and future routes are denoted as $\mathcal{X}^o=\{x_{i}^{o}\}_{i=1}^{|\mathcal{X}|}$ and $\mathcal{X}^f=\{x_{i}^{f}\}_{i=1}^{|\mathcal{X}|}$, respectively.
\end{definition}
 
Given the above definitions, the short-term route prediction (or route prediction for short) problem can be broadly defined as the task of predicting the future route based on observed routes. However, as discussed in Section \ref{sec:intro}, the availability of goal information plays a pivotal role in routing tasks. In some scenarios, no goal information is available. In other cases, we may know the rough direction of the destination, or its exact location.
The degree of goal information inclusion can greatly influence the specific formulation of route prediction \cite{dendorfer2020goal}. Remarkably, no existing work has undertaken an exhaustive evaluation encompassing all these distinct scenarios. Consequently, in this study, we categorize the route prediction problem into three subproblems:

\begin{problem}[Route Prediction $\mathcal{F}$]
Generally, the route prediction problem aims to learn a function $\mathcal{F}$ that maps observed routes to future routes. We identify three distinct subproblems that arise based on the availability of the goal information:

\textup{\textbf{Subproblem 1} (Route prediction with unknown goal $\mathcal{F}_1$)}
The goal information is completely absent from the input. The mapping function $\mathcal{F}_1$ is designed to predict the future routes solely based on the observed routes, disregarding any goal information:
\begin{equation} \label{eq:1}
    \left[\left\{x^{o}_{i}\right\}_{i=1}^{|\mathcal{X}|};\mathbf{G}\right]\stackrel{\mathcal{F}_1(\cdot;\Theta_1)}{\longrightarrow}\{x^{f}_{i}\}_{i=1}^{|\mathcal{X}|},
\end{equation}

\textup{\textbf{Subproblem 2} (Route prediction with goal direction only $\mathcal{F}_2$)}  
The goal direction $r^d_i$ (the relative orientation from the last observed road link to the goal link on the road network) is known in addition to the observed routes. The mapping function $\mathcal{F}_2$ leverages the goal direction to predict the future routes more accurately:
\begin{equation} \label{eq:2}
    \left[\left\{x^{o}_{i}; r^d_i\right\}_{i=1}^{|\mathcal{X}|};\mathbf{G}\right]\stackrel{\mathcal{F}_2(\cdot;\Theta_2)}{\longrightarrow}\{x^{f}_{i}\}_{i=1}^{|\mathcal{X}|},
\end{equation}

\textup{\textbf{Subproblem 3} (Route prediction with complete goal information $\mathcal{F}_3$)}  
Complete goal information is given in the input. The mapping function $\mathcal{F}_3$ leverages both the goal direction $r^d_i$ and exact goal link $e^{\Gamma + \Gamma^\prime}_{i}$ to generate more accurate predictions of the future routes:
\begin{equation} \label{eq:3}
    \left[\left\{x^{o}_{i}; r^d_i; e^{\Gamma + \Gamma^\prime}_{i}\right\}_{i=1}^{|\mathcal{X}|};\mathbf{G}\right]\stackrel{\mathcal{F}_3(\cdot;\Theta_3)}{\longrightarrow}\{x^{f}_{i}\}_{i=1}^{|\mathcal{X}|},
\end{equation}
where $x_i^o$ is the $i$-th observed route, and $\Theta_1, \Theta_2, \Theta_3$ are the parameter sets of the mapping functions $\mathcal{F}_1, \mathcal{F}_2, \mathcal{F}_3$.
\end{problem}

In the context of routing applications, it is crucial to account for various destination-specific requirements. By addressing the routing prediction problem through the three identified subproblems, our study offers valuable empirical evidence regarding the impact of different degrees of goal information availability in real-world scenarios.

\subsection{Knowledge Graph} \label{subsec:pre:kg}

A KG is a heterogeneous structured data representation containing entities (nodes) and their interrelations (edges). The edges carry precise semantic information about the relation type or associated attributes. Formally, the graph is often represented by triplets: $\mathcal{G}=\left\{(h, r, t) \text{ } | \text{ } h, t \in \mathcal{E}, r \in \mathcal{R}\right\}$, where $h$ represents the head entity, $r$ the relation, and $t$ the tail entity. $\mathcal{E}$ is the set of entities, and $\mathcal{R}$ the set of relations. These triplets concisely encode factual information for efficient knowledge discovery, inference, and integration. 
The graph not only serves as a repository of existing knowledge but also facilitates the inference of missing information. This process, known as Knowledge Graph Completion (KGC), finds a tail entity $\hat{t}$ given a head entity and a relation, denoted as $(h, r, \hat{t})$, or its reverse, denoted as $(\hat{h}, r, t)$, completing a partial triplet.

To enhance KGC and provide quantitative measures of relations, KG embedding maps entities and relations to a low-dimensional space, preserving the relational structure. The embedding process can be formalized as two mapping functions $\mathcal{M}_{\mathcal{E}}: \mathcal{E} \rightarrow \mathbb{R}^{\delta_{\mathcal{E}}}$ and $\mathcal{M}_{\mathcal{R}}: \mathcal{R} \rightarrow \mathbb{R}^{\delta_{\mathcal{R}}}$, where ${\delta_{\mathcal{E}}}$ and ${\delta_{\mathcal{R}}}$ are the dimensions of the entity embedding space and relation embedding space.
A scoring function $\phi: \mathbb{R}^{\delta_{\mathcal{E}}} \times \mathbb{R}^{\delta_{\mathcal{R}}} \times \mathbb{R}^{\delta_{\mathcal{E}}} \rightarrow \mathbb{R}$ computes the plausibility of a relation $r$ between entities $h$ and $t$ in the embedded space. The function is defined such that $\phi(\mathcal{M}_{\mathcal{E}}(h), \mathcal{M}_{\mathcal{R}}(r), \mathcal{M}_{\mathcal{E}}(t))$ returns a real number representing the score of the triplet $(h, r, t)$.
KGC infers missing relations or entities by identifying triplets with high scores under the scoring function. This embedding mechanism, coupled with a scoring function, computes and extends the encoded relations within the KG, providing a robust knowledge discovery and integration tool.

\section{Methodology} \label{sec:method}

\subsection{RouteKG Framework Overview}
This section introduces RouteKG, the proposed solution to the route prediction problem. As depicted in Figure~\ref{fig:flow}, the model comprises four modules, namely \textit{Data Preprocessing Module} $\mathcal{M}_{d}$, \textit{Knowledge Graph Module} $\mathcal{M}_{kg}$, \textit{Route Generation Module} $\mathcal{M}_{g}$, and \textit{Rank Refinement Module} $\mathcal{M}_{r}$, each serving a specific purpose and collectively working towards an effective solution.

We start by processing the raw GPS trajectories $\mathcal{T}$ and the raw road network data $G$ with the \textit{Data Preprocessing Module}. This module generates the direction label matrix $\mathbf{D}$, the node adjacency edges (NAE) matrix $\mathbf{A}$, and map-matched routes $\mathcal{X}$. We then divide $\mathcal{X}$ into the observed routes $\mathcal{X}^o$ and future routes $\mathcal{X}^f$. We represent the road network as a MultiDiGraph $\mathbf{G}$ and express the preprocessing step as $\left(\mathcal{X}^o,\mathcal{X}^f,\mathcal{X}, \mathbf{D}, \mathbf{A}\right)=\mathcal{M}_{d}\left(\mathcal{T}, \mathbf{G}\right)$.

Next, the \textit{Knowledge Graph Module} is the core component of the proposed model, it takes the observed routes $\mathcal{X}^o$ and road network $\mathbf{G}$ to predict future routes. It constructs a KG on the road network $\mathbf{G}$, learns spatial relations $\mathcal{R}$, and predicts future routes, converting $\mathcal{X}^o$ to future route probabilities $\mathrm{Pr}(\widetilde{\mathcal{X}^f})$ via $\mathrm{Pr}(\widetilde{\mathcal{X}^f}), \mathcal{R}=\mathcal{M}_{kg}\left(\mathcal{X}^o, \mathbf{G}, \mathbf{D};\Theta_{kg}\right)$, where $\Theta_{kg}$ are the module's parameters, and $\mathcal{R}$ represents the learned spatial relations.

With the estimated future route probabilities $\mathrm{Pr}(\widetilde{\mathcal{X}^f})$, the \textit{Route Generation Module} employs an $n$-ary tree algorithm to generate potential future routes, yielding the top-$K$ preliminary route predictions. This step can be captured by $\{\widetilde{\mathcal{X}^f_k}\}_{k=1}^{K}=\mathcal{M}_{g}(\mathrm{Pr}(\widetilde{\mathcal{X}^f}), \mathbf{G}, \mathbf{A})$, where $\widetilde{\mathcal{X}^f_k}$ denotes the $k$-th generated future route.

In future route prediction, predicted road links at different time steps are not independent but related. Thus, the \textit{Rank Refinement Module} is designed to collectively learn and assess the predicted route. It takes the initial top-$K$ predictions, $\{\widetilde{\mathcal{X}^f_k}\}_{k=1}^{K}$, and refines them using the spatial relations, $\mathcal{R}$, through the mapping $\{\widehat{\mathcal{X}^f_k}\}_{k=1}^{K}=\mathcal{M}_{r}(\{\widetilde{\mathcal{X}^f_k}\}_{k=1}^{K}, \mathcal{R};\Theta_{r})$, where $\Theta_r$ are the module's parameters. This stage ensures that the final route predictions are accurate by considering the sequence of routes and spatial relations.

The motivations and details of the four modules will be explained in the following subsections.

\subsection{Data Preprocessing Module}
To facilitate the KG-related process and route prediction, we first need to perform specific calculations on the road network. This subsection details the method for producing the necessary data for the model components, which aims to compute routes $\mathcal{X}$, route directions $\mathcal{X}_d$, link-to-link direction matrix $\mathbf{D} \in \mathbb{R}^{|\mathbf{E}| \times |\mathbf{E}|}$, and node adjacency edges matrix $\mathbf{A} \in \mathbb{R}^{|\mathbf{V}| \times N_{A}}$,  where the $N_{A}$ is the maximum number of the adjacent edges of all nodes in the $\mathbf{G}$.

\begin{figure*}[ht]
    \centering
    \includegraphics[width=.7\textwidth]{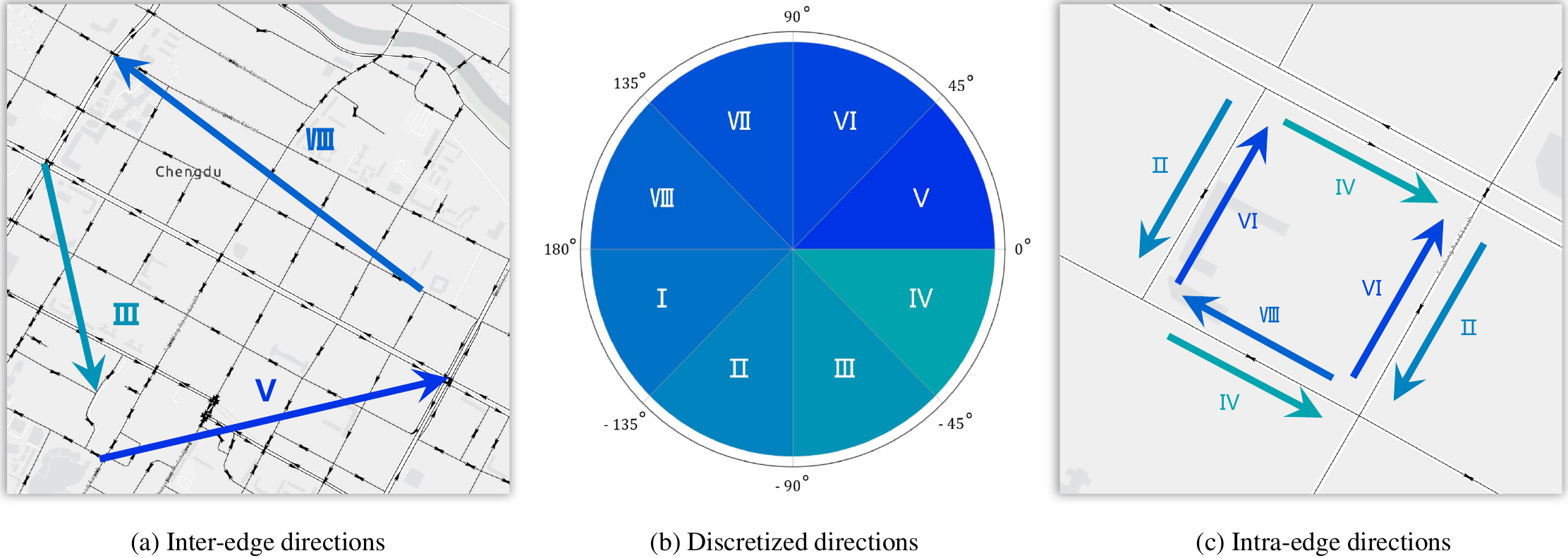}
    \caption{An illustration of discretized directions with examples of inter- and intra-edge directions.}
    \label{fig:direction}
    \vspace{-3mm}
\end{figure*}

Routes $\mathcal{X}$ are obtained by map-matching GPS trajectories $\mathcal{T}$ to the road network $\mathbf{G}$ \cite{Yang2018FastMM}. These routes are then divided into observed routes $\mathcal{X}^o$ and future routes $\mathcal{X}^f$. Considering the importance of direction information in navigation \cite{chrastil2015active}, we discretize continuous directions into $N_d$ classes to form $\mathcal{X}_d$ and $\mathbf{D}$. Figure \ref{fig:direction} provides an example based on $N_d=8$. It has been shown that 8 directions are adequate in uniquely mapping most link-to-link movements and can enhance route prediction performance \cite{liang2021nettraj}. This discretization allows for the convenient computation and assignment of inter- and intra-edge direction labels. It is worth noting that the two-way roads are given only one direction label for simplicity.

To preserve the road network structure information, we construct the node adjacency edges (NAE) matrix, denoted as $\mathbf{A}$, which can be derived directly from the road network $\mathbf{G}$. To build $\mathbf{A}$, we pad the edges adjacent to each node to a uniform length, thereby creating a matrix of dimensions $\mathbb{R}^{|\mathbf{V}| \times N_{A}}$. In this context, $|\mathbf{V}|$ indicates the total number of nodes in the road network, while $N_{A}$ represents the maximum number of adjacent edges to any node. This padding approach enables batch training combined with smart masking techniques.
This matrix not only encodes structural adjacency but also enables efficient batch retrieval of neighboring edge attributes (including embeddings), which is critical for learning relation-aware representations and route reranking.

\subsection{Knowledge Graph Module}

After data preprocessing, we designed a \textit{Knowledge Graph Module} that adapts the KG to the road network, which learns the complex spatial relationships between road links and therefore more accurately estimates the probability of each link as part of the future route $x^f$, given an observed route $x^o$. Formally, given a road network $\mathbf{G}$ and an observed route $x^o \in \mathcal{X}^o$, the module outputs $\Gamma^\prime$ probability distributions $\mathrm{Pr}(\widetilde{\mathcal{X}^f})=\left\{\mathrm{Pr}(\widetilde{x^{f, \gamma}})\right\}_{\gamma=1}^{\Gamma^\prime}$. Each distribution indicates the probability of a link being part of future routes, with the $\gamma$-th distribution indicating the likelihood of each road link being the $\gamma$-th link in those future routes, where $\gamma=1, 2, \dots, \Gamma^\prime$.

Intuitively, a user's route choice is based on their intended goal. Therefore, using KGC for route prediction aligns with the logic behind drivers' route selections. However, most existing KGs are designed for search engines \cite{xiong2017explicit} and text-based Question Answering \cite{huang2019knowledge}, making them unsuitable for direct application to road networks. Therefore, we need to construct a KG tailored to the characteristics of road networks, redefine the KGC problem in this context, and use learned spatial relations for more accurate route prediction. These tasks are encompassed in three submodules we've designed: \textit{Knowledge Graph Construction}, \textit{Knowledge Graph Representation Learning}, and \textit{Future Route Prediction through KGC}. We will detail these in the following subsections.

\subsubsection{Knowledge Graph Construction} \label{subsubsec:spkgconstruct}

To design a KG $\mathcal{G}$ tailored for road networks and route prediction, we first need to select the crucial spatial and structural features in road networks. The desired KG should preserve the spatial relations amongst the identified entities while maintaining its applicability and generalizability across fine-grained scenarios on road networks. In alignment with this objective, the selection focuses solely on those entities and relations that pervade all road networks and routing contexts. A detailed explanation of the entity and relation selection processes is provided below.

\paragraph{\textbf{Entity selection}}
When constructing the KG $\mathcal{G}$ for road networks and routes, the initial step is to identify entities $\mathcal{E}$. In the context of a road network, the predominant entity is the link. Each link $e$ is characterized by its unique identifiers and associated attributes such as length or connectivity. Selecting links as the sole entities reflects their intrinsic importance within the road network. It also ensures the generalizability of the proposed approach in various contexts and scenarios.

\paragraph{\textbf{Relation selection}}

As discussed earlier, route prediction is reformulated as a KGC problem. Based on the selected entities ({\em i.e.,} links), the relations $\mathcal{R}$ chosen for the road network should reflect and preserve the following features: (1) the spatial and structural properties of the road network, and (2) the consistency and preference patterns in observed route selections. Given this, we identify four relations to construct the KG, as illustrated in Table~\ref{tab:relation}.

\renewcommand{\arraystretch}{1.1}
\begin{table}[h]
\centering

\caption{Major spatial relation types, corresponding notations and data sources.}
\resizebox{\linewidth}{!}{
\begin{tabular}{lll}
\toprule 
Relation       & Notation & Data Source \\ \midrule 
ConnectBy      & $\mathcal{R}^c$         &  Road Network $\mathbf{G}$  \\ 
ConsistentWith & $\mathcal{R}^s$         &  Road Network $\mathbf{G}$, Observed routes $\mathcal{X}^o$  \\ 
DistanceTo     & $\mathcal{R}^a$         &  Road Network $\mathbf{G}$, Observed routes $\mathcal{X}^o$  \\ 
DirectionTo    & $\mathcal{R}^d$         &  Road Network $\mathbf{G}$  \\ \bottomrule
\end{tabular}
}
\label{tab:relation}
\end{table}

\vspace{-5pt}

The \emph{ConnectBy} relation, denoted as $\mathcal{R}_c$, defines whether two links are directly connected via a shared node (intersection). Mathematically, for links $e^i$ and $e^j$, this relation is represented as: $(e^i, \mathcal{R}_c, e^j) = \begin{cases} 
1, & \text{if } e^i \text{ and } e^j \text{ share a node} \\
0, & \text{otherwise}
\end{cases}$. This relation models the physical connectivity between links, helping the model understand the road network’s layout.

The \emph{ConsistentWith} relation, denoted as $\mathcal{R}_s$, captures the co-occurrence of two links in the same observed routes. A higher co-occurrence rate indicates a stronger \emph{ConsistentWith} relation. This relation is defined as:
$(e^i, \mathcal{R}_s, e^j) = \frac{|\{ x \in \mathcal{X}^o : e^i \in x, e^j \in x \}|}{|\mathcal{X}^o|},$ where $\mathcal{X}^o$ is the set of observed routes. This reflects driver behavior, indicating which links are commonly used together in routes, and helps capture real-world routing patterns.

The \emph{DistanceTo} relation, denoted as $\mathcal{R}_a$, measures the physical distance between two links. For two links $e^i$ and $e^j$, it is defined as the Euclidean or shortest-path distance: $(e^i, \mathcal{R}_a, e^j) = d(e^i, e^j),$ where $d(\cdot, \cdot)$ represents a distance metric, which can be physical distance or network distance. This relation reflects the spatial proximity of links and helps the model understand how distance affects route selection.

The \emph{DirectionTo} relation, denoted as $\mathcal{R}_d$, represents the navigational direction between two links. For two links $e^i$ and $e^j$, it is defined as a discrete direction class:
$(e^i, \mathcal{R}_d, e^j) = dir(e^i, e^j),$ where $dir(e^i, e^j)$ represents the discretized direction class. This relation encodes navigational decisions, such as turns or straight paths, which are crucial for route planning.

Note that each relation type may contain multiple relations. For example, the relation type ``DirectionTo'' contains $N_d$ directions, indicating a total of $N_d$ direction relations. By comprehensively capturing these four types of relations, the KG offers a rich and nuanced representation of the road network, which can facilitate various routing tasks.

\subsubsection{Knowledge Graph Representation Learning}

Spatial relations between entities ({\em i.e.} links) on a road network should be route-agnostic. This means that these relations should be independent of specific routes and instead solely reflect the spatial attributes of the road network itself. These relations also need to be encoded efficiently to support training and inference. One common approach is to employ KG embedding techniques, which aim to find embedding functions $\mathcal{M}_{\mathcal{E}}$, $\mathcal{M}_{\mathcal{R}}$ that map each entity and each relation into a feature vector. The embedding function $\mathcal{M}_{\mathcal{E}}(\cdot)$ and $\mathcal{M}_{\mathcal{R}}(\cdot)$ should preserve the inherent property of $\mathcal{G}$. However, road networks exhibit complex relations, as illustrated in Figure \ref{fig:relation}, involving many-to-many relations. To address this complexity, we modify and adapt the translation distance model TransH \cite{bordes2013translating} in our study, so that we can effectively learn the vector representations of both entities and relations in $\mathcal{G}$.

\begin{figure*}[h]
    \setcounter{figure}{2}
    \centering
    \includegraphics[width=0.95\textwidth]{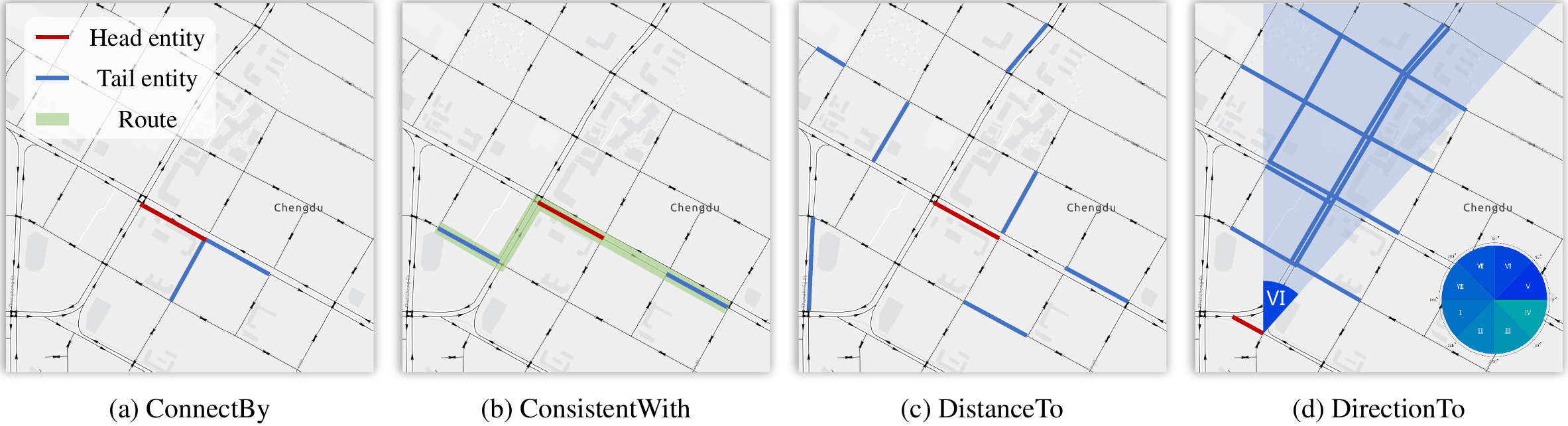}
    \caption{An illustration of many-to-many spatial relations between sampled \emph{head} and \emph{tail} entities. Each relation captures specific patterns between entity pairs: (a) \emph{ConnectBy}: adjacency in the road network via a shared node; (b) \emph{ConsistentWith}: co-occurrence in observed routes; (c) \emph{DistanceTo}: spatial or network distance; (d) \emph{DirectionTo}: navigational direction.}

    \label{fig:relation}
    \vspace{-3mm}
\end{figure*}

To enable KG representation learning, we first need to construct sets of positive triplets $\Delta$ and negative triplets $\Delta^\prime$ for each relation type. Positive triplets reflect connections (e.g., physically connected links or frequently co-used links), while negative triplets are generated by sampling incorrect relationships. To facilitate batch representation learning and ensure comprehensive learning of all entities and relations, we employ a random sampling-based method. The specific construction process for each relation type is described below.

\paragraph{\textbf{ConnectBy} $\mathcal{R}^c$}
For $\mathcal{R}^c$, positive triplets $\Delta_{\mathcal{R}^c}$ are sampled from adjacent edges in spatial graph $\mathbf{G}$, linked by ``ConnectBy''. Negative triplets $\Delta^\prime_{\mathcal{R}^c}$, conversely, are sampled from non-adjacent edges.

\paragraph{\textbf{ConsistentWith} $\mathcal{R}^s$}
$\mathcal{R}^s$ captures transition patterns between links that frequently co-occur within the same observed route, indicating a form of spatial or behavioral consistency. To construct the positive set $\Delta_{\mathcal{R}^s}$ and negative set $\Delta^\prime_{\mathcal{R}^s}$, we utilize the spatial graph $\mathbf{G}$ and observed routes $\mathcal{X}^o$. The positive set $\Delta_{\mathcal{R}^s}$ contains pairs of links that co-occur in the same observed route. To construct the negative set $\Delta^\prime_{\mathcal{R}^s}$, we randomly sample link pairs from $\mathbf{G}$ that do not co-occur in any observed route in $\mathcal{X}^o$. While we do not explicitly encode co-occurrence frequency, frequent entity pairs are more likely to be sampled, which implicitly reflects the defined $\mathcal{R}^s$ relation.

\paragraph{\textbf{DistanceTo} $\mathcal{R}^a$}
For $\mathcal{R}^a$, sampling is performed using observed routes $\mathcal{X}^o$, better aligning the ``DistanceTo'' relation with route prediction and reducing the possible $\mathcal{R}^a$ relations. The resulting sets are $\Delta_{\mathcal{R}^a}$ and $\Delta^\prime_{\mathcal{R}^a}$ for positive and negative triplets, respectively.

\paragraph{\textbf{DirectionTo} $\mathcal{R}^d$}
We need the inter-link direction matrix $\mathbf{D}$ to construct positive and negative triplet sets $\Delta_{\mathcal{R}^d}$ and $\Delta^\prime_{\mathcal{R}^d}$. For sampled edges $e^i,e^j \in \mathbf{E}$, their relative direction $r^d$ is given by $r^d = \mathbf{D}_{e^ie^j}$, forming the positive set. The negative set is formed using edge pairs that contradict the directional relation in $\mathbf{D}$ (i.e., opposite direction).

We denote the positive triplet sets for all relations as $\Delta_{\mathcal{R}^\cdot}$ and the negative sets as $\Delta_{\mathcal{R}^\cdot}^\prime$. Consider an identified type of relation $\mathcal{R}^{\cdot}$, and we incorporate two trainable weight matrices. One matrix functions as the relation embedding matrix, represented as $\mathbf{W}_{\mathcal{R}^{\cdot}} \in \mathbb{R}^{|\mathcal{R}^{\cdot}| \times \delta_{\mathcal{R}^{\cdot}}}$, while the other corresponds to the relation hyperplane, denoted as $\mathbf{P}_{\mathcal{R}^{\cdot}}$, both maintaining congruent dimensions. The relation hyperplane $\mathbf{P}_{\mathcal{R}^{\cdot}}$ defines a learned subspace onto which entity embeddings are projected, enabling the model to capture relational constraints in a more structured and discriminative manner.
Given the sets of positive triplets $\Delta_{\cdot}$ and negative triplets $\Delta_{\cdot}^{\prime}$, where $\cdot$ can represent any of the relations on the KG, the representation learning process involves the following steps. For any triplet $(h, r, t) \in \Delta_{\cdot}$ and $(h^\prime, r^\prime, t^\prime) \in \Delta^\prime_{\cdot}$, we use $\mathbf{h}$, $\mathbf{r}$, $\mathbf{t}$, $\mathbf{h}^\prime$, $\mathbf{r}^\prime$, and $\mathbf{t}^\prime$ denoted their embeddings and use $\mathbf{p}^r$ and $\mathbf{p}^{r^\prime}$ to denote the hyperplane of relation $r$ and $r^\prime$ with $\left(\mathbf{p}^r\right)^\top$ and $\left(\mathbf{p}^{r^\prime}\right)^\top$ as their transposes. For all relations, the loss function for learning the representation of the KG $\mathcal{G}$ is:
\begin{equation} \label{eq:5}
    \begin{aligned}
        &\mathcal{L}_{rep}=\sum_{\Delta, \Delta^\prime \in \left\{(\Delta_{\mathcal{R}^\cdot}, \Delta_{\mathcal{R}^\cdot}^\prime)\right\}}\sum_{(h,r,t) \in \Delta} \sum_{(h^\prime,r^\prime,t^\prime) \in \Delta^\prime} \\
        & \Biggl[\left\|\text{ }\left(\mathbf{h}-\left(\mathbf{p}^r\right)^\top \mathbf{h} \mathbf{p}^r\right) + \mathbf{r} - \left(\mathbf{t}-\left(\mathbf{p}^r\right)^\top \mathbf{t} \mathbf{p}^r\right)\text{ }\right\|_{\ell_1}  + \psi - \\
        &\left\|\text{ }\left(\mathbf{h}^\prime-(\mathbf{p}^{r^\prime})^\top \mathbf{h}^\prime \mathbf{p}^{r^\prime}\right) + \mathbf{r}^\prime - \left(\mathbf{t}^\prime-(\mathbf{p}^{r^\prime})^\top \mathbf{t}^\prime \mathbf{p}^{r^\prime}\right)\text{ }\right\|_{\ell_1}\text{ }\Biggr]_{+},
    \end{aligned}
\end{equation}

Eq. \eqref{eq:5} defines the margin loss $\mathcal{L}_{rep}$, which calculates the difference in scores between positive and negative triplets. The margin $\psi$ ensures a separation between the scores of positive and negative triplets. Before each batch training starts, we impose a constraint to ensure that $\textbf{p}^r$ and $\textbf{p}^{r^\prime}$ are unit normal vectors by projecting them to the unit $\ell_2$-ball: $\forall r \in \mathcal{R}, \left\|\mathbf{p}^r\right\|_2=1$.

\subsubsection{Future Route Prediction through Knowledge Graph Completion}

\begin{figure}[ht]
    \centering
    \includegraphics[width=.48\textwidth]{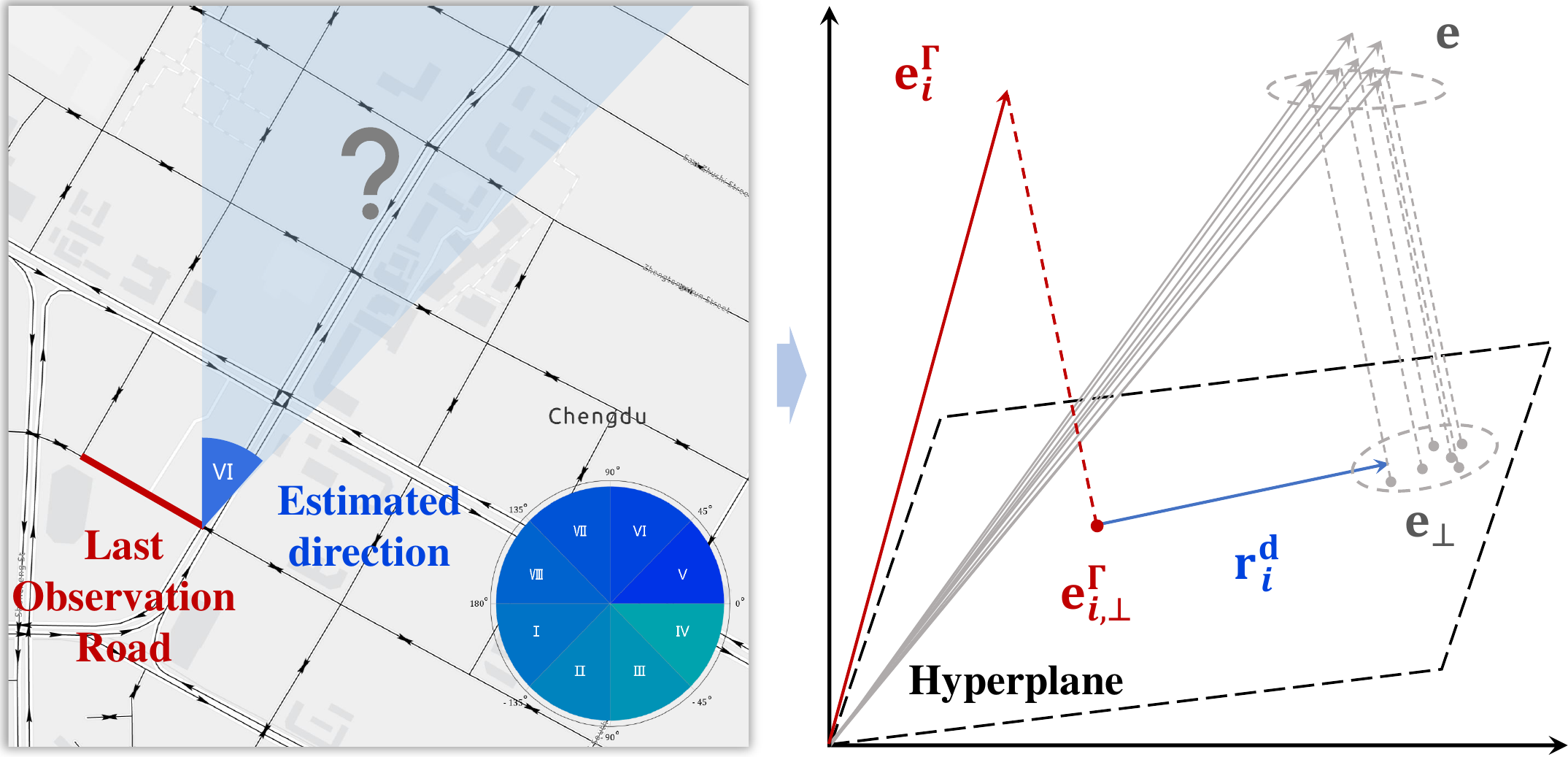}
    \caption{An illustration of knowledge graph completion for future route prediction.}
    \vspace{-3mm}
    \label{fig:kgc}
\end{figure}

Our study approaches the task of future route prediction by framing it as a KGC problem. As shown in Figure \ref{fig:kgc}, given the last link $e_i^{\Gamma}$ ({\em i.e.}, head entity) of the $i$-th observed route $x^o_i$ and the (estimated or actual) direction of movement ({\em i.e.}, relation), our objective is to infer the future route $\widehat{x^f_i}$ ({\em i.e.}, tail entity) that the user will traverse. In our case, the actual direction $r^d_{i}$ of a user's movement is the direction from the current link to the last link of the future route. 
Utilizing KGC, we introduce an innovative objective to predict the immediate future routes of road users. This addition not only utilizes the learned KG embeddings to enhance route prediction accuracy but also enriches the KG representation with deeper semantics, thereby creating a synergistic effect between KG embedding and route prediction.

We use the EGoalD for the whole process illustration and then provide the details for the GoalD and Goal settings.

\paragraph{EGoalD Setting}

We consider the $i$-th observed route $x^o_i \in \mathcal{X}^o$ and its corresponding direction sequence $x^{o, d}_i = \left\{e^{d, j}_i\right\}_{j=\Gamma+1}^{\Gamma+\Gamma^\prime} \in \mathcal{X}^{o,d}$, where $x^o_i=\left\{e^j_i\right\}^{\Gamma}_{j=1}$ represents a sequence of links and $e^{d,j}_i$ denotes the direction label of the $j$-th link. Here, $\mathcal{X}^{o,d}$ denotes the set of all such direction sequences. Initially, we extract the embeddings of all elements of $x^o$ and $x^{o, d}$ by multiplying their respective one-hot vectors with the corresponding trainable embedding matrices: $\mathbf{W}_{\mathcal{E}} \in \mathbb{R}^{|\mathcal{E}| \times \delta_{\mathcal{E}}}$ and $\mathbf{W}_{\mathcal{R}^{d}} \in \mathbb{R}^{|\mathcal{R}^d| \times \delta_{\mathcal{R}^d}}$. The resulting embeddings for $x^{o}_{i}$ and $x^{o, d}_{i}$ are denoted as $\mathbf{x}^{o}_{i}=\{\mathbf{e}^{j}_{i} \Vert\}_{j=1}^{\Gamma}$ and $\mathbf{x}^{o, d}_{i}=\{\mathbf{e}^{d, j}_{i} \Vert \}_{j=\Gamma+1}^{\Gamma+\Gamma^\prime}$, with $\mathbf{e}_{i}^{\cdot} \in \mathbb{R}^{\delta_\mathcal{E}}$, $\mathbf{e}^{d, \cdot}_{i} \in \mathbb{R}^{\delta_{\mathcal{R}^d}}$.

To predict the direction of the user's future routes, we utilize a Multi-layer Perceptron (MLP) \cite{popescu2009multilayer} to encode $\mathbf{x}^{o}_i$ and $\mathbf{x}^{o, d}_i$:
\begin{equation} \label{eq:6}
\widehat{r^{d}_{i}}=\arg\max \left(\textsc{MLP}_d\left(\mathbf{x}^{o}_{i} \text{ } \Vert \text{ } \mathbf{x}^{o, d}_{i}\right)\right),
\end{equation} 
where $\Vert$ is the concatenate operation, $\widehat{r^{d}_{i}}$ represents the estimated direction of the $i$-th future route, and we employ the cross-entropy loss to optimize the parameters of the $\mathrm{MLP}_d$:
\begin{equation}\label{eq:7}
    \mathcal{L}_{d}=-\log\mathrm{Softmax}\left(\mathrm{MLP}_d\left({x}^{o}_{i} \text{ } \Vert \text{ } \mathbf{x}^{o, d}_{i}\right)\right)\left[r^d_i\right]
\end{equation}

To predict future routes, the last link $e_i^{\Gamma}$ is converted into the corresponding entity embedding by multiplying the one-hot vector of $e_i^{\Gamma}$ with $\mathbf{W}_{\mathcal{E}}$, yielding $\mathbf{e}_i^{\Gamma} \in \mathbb{R}^{\delta_{\mathcal{E}}}$. Then, $\mathbf{r}^d_i \in \mathbb{R}^{\delta_{\mathcal{R}^{d}}}$ and $\mathbf{p}^{d}_i \in \mathbb{R}^{\delta_{\mathcal{R}^{d}}}$ are obtained through a similar operation with $\mathbf{W}_{\mathcal{R}^d}$ and $\mathbf{P}_{\mathcal{R}^d}$, respectively. Note that in this stage, RouteKG only needs the user's current position ({\em i.e.}, the last link of the observed route), which is fundamentally different from existing seq2seq methods. Given $\mathbf{p}^d_i$, we first project the $\mathbf{e}_i^{\Gamma} \in \mathbb{R}^{\delta_{\mathcal{E}}}$  and all links embeddings $\mathbf{e} \in \mathbb{R}^{|\mathcal{E}| \times \delta_{\mathcal{E}}}$ to the hyperplane $\mathbf{p}^d_i$ to obtain the projected head embedding $\mathbf{e}_{i, \perp}^\Gamma$ and all candidate tail embedding $\mathbf{e}_{\perp}$ on the hyperplane:
\begin{equation}\label{eq:8}
    \begin{aligned}
        \mathbf{e}_{i, \perp}^\Gamma &=\mathbf{e}^\Gamma_i - {\left(\mathbf{p}^d_i\right)}^\top\mathbf{e}^\Gamma_i{\mathbf{p}^d_i}, \\
        \mathbf{e}_{\perp} &=\mathbf{e} - {\left(\mathbf{p}^d_i\right)}^\top\mathbf{e} \text{ }{\mathbf{p}^d_i}, \\
    \end{aligned}
\end{equation}
where $\mathbf{e}_{i, \perp}^\Gamma \in \mathbb{R}^{\delta_{\mathcal{E}}}$ and $\mathbf{e}_{\perp} \in \mathbb{R}^{|\mathcal{E}| \times \delta_{\mathcal{E}}}$. The projection ensures that the similarity computation is constrained to a direction-specific subspace, allowing the model to capture semantics aligned with the intended travel direction.

Upon acquiring the projected head embedding $\mathbf{e}_{i, \perp}^\Gamma$, we add the relation to the projected head embedding to query the tail entity. Given the projected head embedding $\mathbf{e}_{i, \perp}^\Gamma$, the direction relation embedding of the estimated direction $\widehat{\mathbf{r}^d_i}$, and the distance relation embedding $\mathbf{r}^{a, \gamma}_i$, we could query the tail entity based on the following equation: 
\begin{equation} \label{eq:9}
    \mathrm{Pr}(\widetilde{x^{f, \gamma}_{i}}) = \mathrm{Softmax}\left(\mathbf{e}_{\perp}\cdot\left[\left( \mathbf{e}_{i, \perp}^\Gamma + \widehat{\mathbf{r}^d_i} \right) \odot \mathbf{r}^{a, \gamma}_i\right]^\top\right),
\end{equation}
where $\mathrm{Pr}(\widetilde{x^{f, \gamma}_{i}}) \in \mathbb{R}^{|\mathcal{E}|}$ is the predicted probability distribution which indicates the likelihood of each link being the $\gamma$-th link of the $i$-th future route, $\odot$ denotes element-wise product. We can set $\gamma$ from 1 to $\Gamma^\prime$ and recursively use Eq.~\eqref{eq:7} to obtain $\Gamma^\prime$ probability distributions $\{\mathrm{Pr}(\widetilde{x^{f, \gamma}_i})\}_{\gamma=1}^{\Gamma^\prime}$ representing the estimated future route probabilities, which is the final output of the module. The scoring reflects how likely each candidate link is as the next step, by measuring the directional similarity between the translated current position and candidate links, modulated by distance.

To optimize the KG embeddings, the loss of the future route prediction is defined as:
\begin{equation}\label{eq:11}
    \mathcal{L}_{pred}=-
    \sum_{i=1}^{|\mathcal{X}|} 
    \sum_{\gamma=1}^{\Gamma^\prime}
    \log \mathrm{Pr}(\widetilde{x^{f, \gamma}_i})\left[x_i^{f, \gamma}\right],
\end{equation}
where $x_i^{f, \gamma}$ is the actual $\gamma$-th link of the $i$-th future route, and the indexing operation $[\cdot]$ retrieves the predicted log-probability assigned to the ground-truth link, and the loss corresponds to maximizing the likelihood of the correct link via a standard cross-entropy objective. Note that $x_i^{f, \gamma}$ and $e_i^{\Gamma + \gamma}$ indicate the same link in the $i$-th future route.

\paragraph{GoalD Setting}

It should be noted that the estimation of the user's future route direction is only necessary when the goal is unspecified ({\em i.e.}, subproblem 1). Conversely, when the goal direction or the actual goal is provided, one can directly utilize the given goal direction ({\em i.e.}, subproblems 2 and 3). The embedding of the estimated direction $\widehat{\mathbf{r}^d_i}$ in Eq. (\ref{eq:11}) can be replaced by the actual direction relation embedding $\mathbf{r}^d_i$.

\paragraph{Goal Setting}

For route prediction with complete goal information ({\em i.e.}, subproblem 3), we make a subtle change to Eq.~\eqref{eq:9} by simply add the projected the tail entity ({\em i.e.}, goal) embedding $\mathbf{e}_{i, \perp}^{\Gamma + \Gamma^\prime}$ to the head embedding $\mathbf{e}_{i, \perp}^{\Gamma}$. The tail entity quering process could be updated as:
\begin{equation} \label{eq:10}
    \mathrm{Pr}(\widetilde{x^{f, \gamma}_{i}}) = \mathrm{Softmax}(\mathbf{e}_{\perp}\cdot\left[( \mathbf{e}_{i, \perp}^\Gamma + \mathbf{e}_{i, \perp}^{\Gamma + \Gamma^\prime} + \mathbf{r}^d_i) \odot \mathbf{r}^{a, \gamma}_i\right]^\top),
\end{equation}

\subsection{Route Generation Module}

\begin{figure}[h]
    \centering
    \includegraphics[width=.48\textwidth]{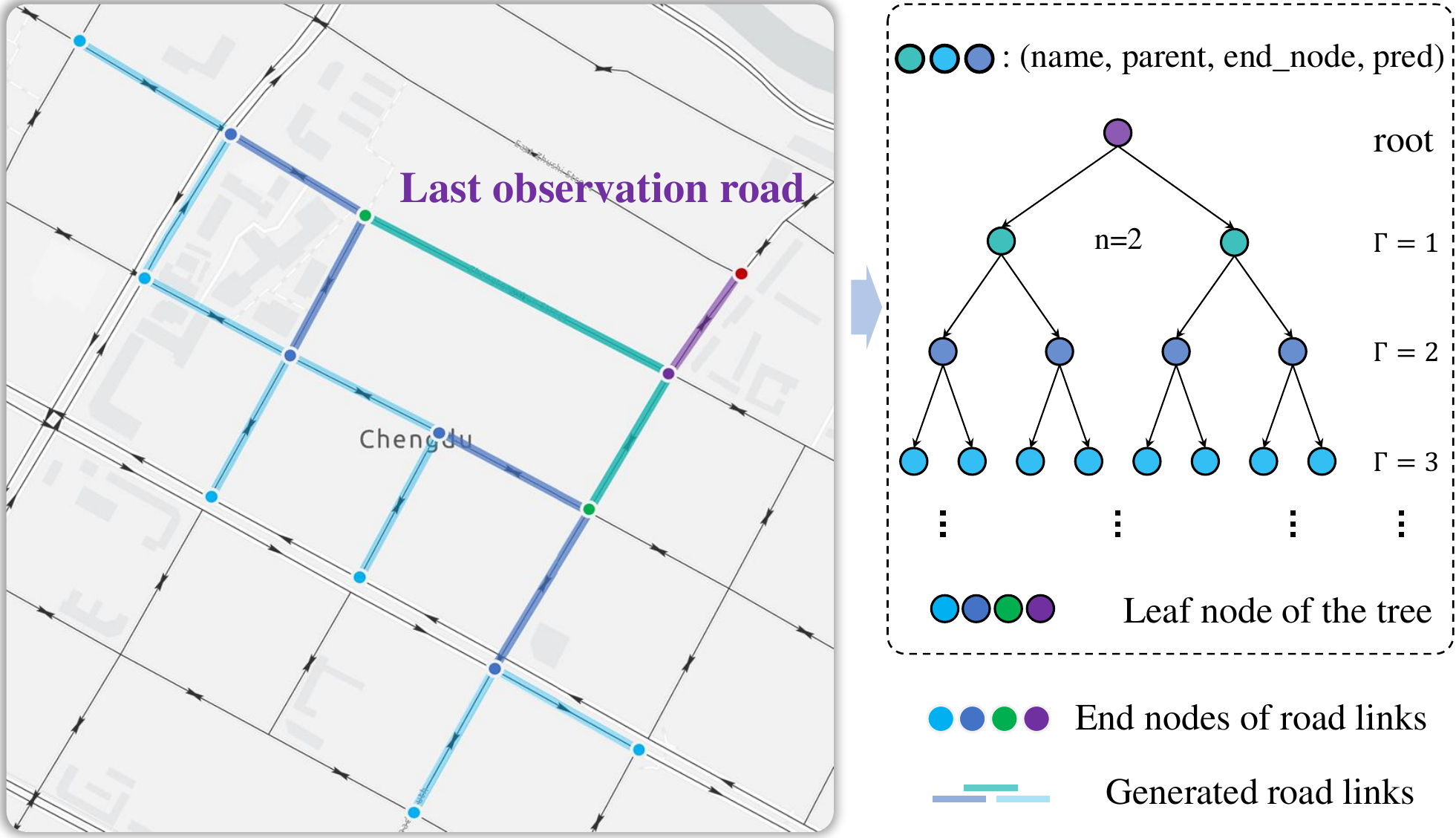}
    \caption{A demonstration of the \textit{Spanning Route} algorithm.}
    
    \label{fig:SpRt}
\end{figure}

Given predicted future route probabilities $\mathrm{Pr}(\widetilde{\mathcal{X}^f})$, the \textit{Route Generation Module} generates multiple possible future routes from these probabilities. An $n$-ary tree-based algorithm, \textit{Spanning Route}, is proposed to generate these future routes based on the predicted probabilities. This algorithm is visualized in Figure \ref{fig:SpRt} through a simplified case where $n=2$ and only $\Gamma=3$ predicted future links are illustrated.
For each tree node, we designate four attributes: \textit{name}, \textit{parent}, \textit{end\_node}, and \textit{pred}. The \textit{name} corresponds to the identification of the leaf, while the \textit{parent} points to the predecessor of the current leaf. The attribute \textit{end\_node} signifies the terminal node of the current predicted link, and \textit{pred} is the present predictions.

\setcounter{algocf}{0}
\begin{algorithm}[h]
\caption{\textbf{Spanning Route}.}
\SetKwInOut{Input}{Input}\SetKwInOut{Output}{Output}
\label{alg:1}

\Input{~$\Gamma^\prime$ probability distributions $\{\mathrm{Pr}(\widetilde{x^{f, \gamma}_i})\}_{\gamma=1}^{\Gamma^\prime}$;
     road network $\mathbf{G}=(\mathbf{V}, \mathbf{E})$; \\
     NAE matrix $\mathbf{A}$; 
     The tree's degree $n$.
    }
\Output{~Top-$K$ predicted future routes $\{\widetilde{x_{i, k}^{f}}\}_{k=1}^{K}$.}
\BlankLine

root $\leftarrow$ \textbf{CreateNewNode}(name = ``root'', parent = NIL, end\_node = $v_{i, \Gamma}^{s}$, pred = NIL) \\
\For{$\gamma=1, \dots, \Gamma^\prime$}{
    leaves $\leftarrow$ \textbf{GetLeaves}(root) \\
    \For{leaf $\in$ leaves}{
        $\mathcal{N}_{end\_node}^{e}$ = $\mathbf{A}[leaf.end\_node, :]$ \\
        $\left\{e_{i, k}^{\Gamma+\gamma}\right\}_{k=1}^{n}$ = \textbf{GetTopK}($\mathrm{Pr}(\widetilde{x^{f, \gamma}_{i}})[\mathcal{N}_{end\_node}^{e}]$, $K=n$) \\
        \For{$k=1, \dots, n$}{
            node = \textbf{CreateNewNode}(name = ``$k$'', parent = leaf, end\_node = $e_{i, k}^{\Gamma+\gamma}[1]$, pred = $e_{i, k}^{\Gamma+\gamma}$)
        }
    }
    
}
leaves $\leftarrow$ \textbf{GetLeaves}(root) \\
\For{$k=1, \dots, K$}{
    $\text{path}_{k}$ = \textbf{Traverse}(root, leaves[$k$]) \\ 
    $\widetilde{x_{i, k}^{f}}$ = $\left\{\text{path}_{k}\text{[i].pred}\right\}_{i=1}^{\Gamma^\prime}$ \\ 
}
\end{algorithm}
\vspace{-5mm}

To formally introduce the \textit{Spanning Route} algorithm, we provide the pseudo-code for generating multiple future routes based on the predicted probabilities in Algorithm \ref{alg:1}. Specifically, the algorithm encompasses four primary functions. The \textbf{CreateNewNode} function instantiates a new node in the tree given its attributes, while the \textbf{GetLeaves} function takes the root node as input and outputs all leaves of the tree. The \textbf{GetTopK} function retrieves the top-$k$ predictions given a predicted probability and $k$, and the \textbf{Traverse} function applies a tree-based Depth-First Search (DFS) traversal algorithm, specifically a Pre-Order traversal \cite{tarjan1972depth}---to acquire the path from the root to a specified leaf. This last function is instrumental in merging the predicted links into cohesive predicted routes. We note that certain detailed masking and indexing operations have been omitted in the presented pseudo-code for clarity. For more details, please refer to the minibatch version of the \textit{Spanning Route} algorithm \ref{alg:srbatch} in the Appendix~\ref{apx:batchalg}.

\subsection{Rank Refinement Module}

The top-$K$ future routes candidates $\widetilde{\mathcal{X}^f}=\{\widetilde{\mathcal{X}^f_k}\}_{k=1}^K=\{\{\widetilde{x_{i, k}^{f}}\}_{k=1}^{K}\}_{i=1}^{|\mathcal{X}^f|}$ offer an initial selection of possible future routes. 
However, the dependencies among different links within these routes are solely based on the connectivity of the road network. Given that the consistency and other spatial relations of links within a route also affect people's choices of routes, a more refined approach is needed for accurate future route prediction. To achieve this, we leverage learned spatial relations $\mathcal{R}$ to model the dependencies between different links and rerank the candidate routes based on the learned dependencies. This process can be denoted as $\{\widehat{\mathcal{X}^f_k}\}_{k=1}^{K}=\mathcal{M}_{r}(\{\widetilde{\mathcal{X}^f_k}\}_{k=1}^{K}, \mathcal{R};\Theta_{r})$ .

Consider the future routes $\widetilde{x_{i}^{f}} \in \mathbb{R}^{K \times \Gamma^\prime}$. Initially, these routes are encoded using embedding matrices $\mathbf{W}_{\mathcal{E}}$ and $\mathbf{W}_{\mathcal{R}^d}$, thereby resulting in the route embedding $\widetilde{\mathbf{x}^{f}_{i}} \in \mathbb{R}^{K \times \Gamma^\prime \times \delta_\mathcal{E}}$ and route direction embedding $\widetilde{\mathbf{x}^{f, d}_{i}} \in \mathbb{R}^{K \times \Gamma^\prime \times \delta_{\mathcal{R}^d}}$. In the subsequent reranking phase, we prioritize the routes with higher consistency and connectivity, utilizing the spatial relations $\mathcal{R}$ learned from the \textit{Knowledge Graph Module}.
Specifically, the obtained route embeddings are projected onto the ``ConnectBy'' and ``ConsistentWith'' hyperplanes as follows:
\begin{equation}\label{eq:12}
    \begin{aligned}
        \widetilde{\mathbf{x}^f_{i, \perp^{c}}} &= \widetilde{\mathbf{x}^f_{i}} - \left(\mathbf{p}^{c}\right)^\top \widetilde{\mathbf{x}^f_{i}} \mathbf{p}^{c} \\
        \widetilde{\mathbf{x}^f_{i, \perp^{s}}} &= \widetilde{\mathbf{x}^f_{i}} - \left(\mathbf{p}^{s}\right)^\top \widetilde{\mathbf{x}^f_{i}} \mathbf{p}^{s}, \\
    \end{aligned}
\end{equation}
where $\widetilde{\mathbf{x}^f_{i, \perp^{c}}} \in \mathbb{R}^{K \times \Gamma^\prime \times \delta_{\mathcal{E}}}$ and $\widetilde{\mathbf{x}^f_{i, \perp^{s}}} \in \mathbb{R}^{K \times \Gamma^\prime \times \delta_{\mathcal{E}}}$ represent the projected route embeddings. 

To quantify the internal consistency and connectivity of the generated routes, related margins for each route are calculated:
\begin{equation}\label{eq:13}
    \begin{aligned}
        \mathbf{r}^{f, c}_{i, m} &= \frac{1}{\Gamma^\prime - 1}\sum_{j=1}^{\Gamma^\prime-1} \widetilde{\mathbf{x}^f_{i, \perp^{c}}}\left[:, j, :\right] - \widetilde{\mathbf{x}^f_{i, \perp^{c}}}\left[:, j+1, :\right] \\
        \mathbf{r}^{f, s}_{i, m} &= \frac{1}{\Gamma^\prime - 1}\sum_{j=1}^{\Gamma^\prime-1} \widetilde{\mathbf{x}^f_{i, \perp^{s}}}\left[:, j, :\right] - \widetilde{\mathbf{x}^f_{i, \perp^{s}}}\left[:, j+1, :\right], \\
    \end{aligned}
\end{equation}
where $\mathbf{r}^{f, c}_{i, m} \in \mathbb{R}^{K \times \delta_{\mathcal{E}}}$ is the connectivity margin and $\mathbf{r}^{f, s}_{i, m} \in \mathbb{R}^{K \times \delta_{\mathcal{E}}}$ the consistency margin. These equations project candidate routes onto relation-specific hyperplanes (e.g., for \emph{ConnectBy} and \emph{ConsistentWith}), allowing the model to assess how well each route follows the learned spatial relations. Routes more aligned with these hyperplanes are considered more plausible and prioritized during reranking. Following this, the derived $\mathbf{r}^{f, c}_{m}$ and $\mathbf{r}^{f, s}_{m}$ are flattened and, together with the flattened route embedding $\widetilde{\mathbf{x}^{f}_{i}} \in \mathbb{R}^{K \cdot \Gamma^\prime \cdot \delta_{\mathcal{E}}}$ and route direction embedding $\widetilde{\mathbf{x}^{f, d}_{i}} \in \mathbb{R}^{K \cdot \Gamma^\prime \cdot \delta_{\mathcal{E}}}$, used to compute the new rank:
\begin{equation}    \label{eq:14}
\mathrm{Pr}(\widetilde{R})=\mathrm{Softmax}(\mathrm{MLP}_{r}(\mathbf{r}^{c}_{m} \text{ } \Vert \text{ } \mathbf{r}^{s}_{m}  \text{ } \Vert  \text{ } \mathrm{MLP}_{f}(\widetilde{\mathbf{x}^{f}_{i}} \text{ } \Vert \text{ } \widetilde{\mathbf{x}^{f, d}_{i}}))),
\end{equation}
where $\mathrm{Pr}(\widetilde{R}) \in \mathbb{R}^{K}$ denotes the probability distribution over the $K$ predicted future routes being the actual future routes. Based on this probability, we can determine the new predicted rank, resulting in reranked future route predictions denoted as $\{\widehat{x^{f}_{i, k}}\}_{k=1}^{K}$. We also adopt the cross-entropy loss for the \textit{Rank Refinement Module}:
\begin{equation}\label{eq:15}
    \mathcal{L}_{rank}=-\log\mathrm{Pr}(\widetilde{R})\left[x^f_i\right], 
\end{equation}

Note that samples from the set of multiple predicted future routes are excluded if they do not contain a ground truth future route.
While our KG facilitates an effective overall selection ({\em i.e.}, top-$K$) of future routes, it is crucial to notice that an enhancement in the top-$K$ predictions does not necessarily translate to a superior top-$1$ prediction. Therefore, we directly refine the top-$1$ prediction using the initial predictions in our implementation. This is done by employing an MLP to encode the initial prediction embeddings $\mathbf{x}^{f}_{i}$ and $\mathbf{x}^{f, d}_{i}$, the last observed link $\mathbf{e}^{\Gamma}_{i}$ and $\mathbf{e}^{d, \Gamma}_{i}$, and estimated goal direction $\mathbf{r}^d_{i}$:
\begin{equation}\label{eq:16}
    \widetilde{x^f_i}=\mathrm{MLP}_{k}(\mathbf{e}^{\Gamma}_{i} \text{ } \Vert \text{ } \mathbf{e}^{d, \Gamma}_{i} \text{ } \Vert \text{ } \mathbf{r}^d_{i} \text{ } \Vert \text{ } \mathrm{MLP}_{x}(\mathbf{x}^{f}_{i} \text{ } \Vert \text{ } \mathbf{x}^{f, d}_{i})),
\end{equation}
where $\widetilde{x^f_i} \in \mathbb{R}^{\Gamma^\prime \times |\mathcal{E}|}$, is also optimized by minimizing the corresponding cross-entropy loss. Subsequently, the top-$1$ prediction is generated using the \textit{Route Generation Module} for the $n=1$ case. The generated future route can be inserted into the first position, replacing the original $K$-th prediction, to obtain the refined top-$K$ future route predictions $\{\widehat{x^{f}_{i, k}}\}_{k=1}^{K}$.

\subsection{Multi-Objectives Optimization}

The objective of RouteKG is to leverage learned spatial relations to predict future routes. This is achieved through a multi-objective learning framework, combining several complementary goals: (1) $\mathcal{L}_{rep}$ encourages high-quality KG embeddings by modeling multiple spatial relations; (2) $\mathcal{L}_{d}$ ensures accurate direction classification for the EGoalD setting; (3) $\mathcal{L}_{pred}$ guides the prediction of future links via KGC; (4) $\mathcal{L}_{rank}$ improves final route quality by reranking based on relational consistency. Collectively, these objectives support each stage of the prediction pipeline, ensuring that the model benefits from both structural knowledge and task-specific supervision.
These objectives are jointly optimized through the following weighted total loss:
\begin{equation}\label{eq:17}
    \mathcal{L}=w_{rep}\cdot\mathcal{L}_{rep}+w_{d}\cdot\mathcal{L}_{d}+w_{pred}\cdot\mathcal{L}_{pred}+w_{rank}\cdot\mathcal{L}_{rank},
\end{equation}
where $w_\cdot$ are the balancing weights. The overall training procedure that integrates all components is summarized in Algorithm~\ref{alg:learning}.

\setcounter{algocf}{1}
\begin{algorithm}[h]
\caption{RouteKG training procedure.}
\label{alg:learning}
\SetKwInOut{Input}{Input}\SetKwInOut{Output}{Output}

\Input{~A batch of observed routes $\mathcal{X}^o_{\mathcal{B}} \in \mathbb{R}^{\mathcal{B} \times \Gamma}$;\\
Road network $\mathbf{G}$; NAE matrix $\mathbf{A}$;\\
Inter-road direction matrix $\mathbf{D}$;\\

}
\BlankLine
\textbf{Init:} Randomly initialize KG parameters $\Theta_{kg}$ and rerank parameters $\Theta_r$.\;
\For{$m = 1,\dots,\textit{max\_iters}$}{
    \textcolor{gray}{// --- Step 1: KG representation learning ---}\\
    Normalize hyperplane embeddings $\left\|\mathbf{P}_{\mathcal{R}^\cdot}\right\|_2=1$\;
    Sample positive/negative triplets 
    $\Delta,\Delta'$ from $\mathcal{X}^o_{\mathcal{B}}$ and $\mathbf{G}$\;
    Compute $\mathcal{L}_{rep}$ (Eq.~4)\;
    \BlankLine
    
    \textcolor{gray}{// --- Step 2: forward prediction ---}\\
    \uIf{\textit{setting} = NoGoal}{
        $\mathrm{Pr}(\widetilde{\mathcal{X}^f}),\mathcal{R},\widehat{r^{d}}\leftarrow\mathcal{M}_{kg}\!\bigl(x^o,\,\mathbf{G},\mathbf{D};\Theta_{kg}\bigr)$\\
        Compute $\mathcal{L}_d$ (Eq.~6)
    }
    \uElseIf{\textit{setting} = GoalD}{
        $\mathrm{Pr}(\widetilde{\mathcal{X}^f}),\mathcal{R}\leftarrow
        \mathcal{M}_{kg}\!\bigl(\{x^o,r^d\},\mathbf{G},\mathbf{D};\Theta_{kg}\bigr)$
    }
    \uElseIf{\textit{setting} = Goal}{
        $\mathrm{Pr}(\widetilde{\mathcal{X}^f}),\mathcal{R}\leftarrow
        \mathcal{M}_{kg}\!\bigl(\{x^o,r^d,e^{\Gamma+\Gamma'}\},\mathbf{G},\mathbf{D};\Theta_{kg}\bigr)$
    }
    \BlankLine
    \textcolor{gray}{// --- Step 3: spanning-route + rerank ---}\\
    Candidate routes $\{\widetilde{\mathcal{X}^f_k}\}_{k=1}^{K}
       \leftarrow \mathcal{M}_{g}\bigl(\mathrm{Pr}(\widetilde{\mathcal{X}^f}),\mathbf{G},\mathbf{A}\bigr)$\;
    Compute $\mathcal{L}_{pred}$ (Eq.~10)\;
    Ranked routes $\{\widehat{\mathcal{X}^f_k}\}_{k=1}^{K}
       \leftarrow \mathcal{M}_{r}(\{\widetilde{\mathcal{X}^f_k}\},\mathcal{R};\Theta_r)$\;
    Compute $\mathcal{L}_{rank}$ (Eq.~14)
    \BlankLine
    \textcolor{gray}{// --- Step 4: parameter update ---}\\
    $\Theta_{kg}\! \leftarrow\! \Theta_{kg} - \eta\nabla_{\Theta_{kg}}
       (\mathcal{L}_{rep}+\mathcal{L}_{d}+\mathcal{L}_{pred})$\;
    $\Theta_{r}\! \leftarrow\! \Theta_{r} - \eta\nabla_{\Theta_{r}}
       (\mathcal{L}_{rank})$\;
}
\end{algorithm}

\section{Experiments}\label{sec:exp}

\subsection{Data}

We conduct experiments on taxi trajectory data obtained from two cities in China: Chengdu and Shanghai. The Chengdu dataset was acquired from the Didi Chuxing GAIA Initiative\footnote{https://gaia.didichuxing.com}. It contains the records of 143,888 drivers, covering a month of data from November 1, 2016, to November 30, 2016, with an average sampling rate of 2~4 seconds. The selected region in Chengdu spans from 30.65°N to 30.73°N in latitude and 104.04°E to 104.13°E in longitude, with the region's road network comprising 2,832 nodes and 6,506 edges. The Chengdu data record incorporates driver ID, order ID, timestamp, longitude, and latitude. This study used the first seven days of Chengdu's data.

The Shanghai dataset consists of trajectory records from 10,609 taxis from April 16, 2015, to April 21, 2015, with an average sampling rate of approximately 10 seconds per record. We concentrated on a specific region in Shanghai, adhering to the parameters outlined by \cite{zhao2023deep}. The chosen region's road network incorporates 320 nodes and 714 links. Each data entry includes the taxi ID, date, time, longitude, latitude, and an occupied flag indicator.

\begin{figure}[ht]
    \centering
    \includegraphics[width=.48\textwidth]{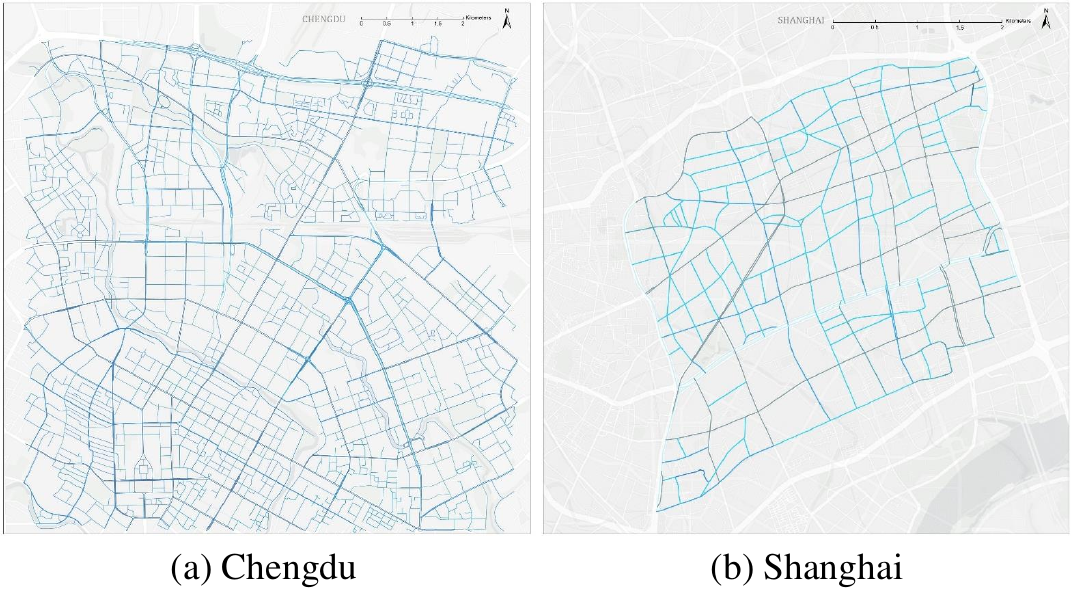}
    \caption{Selected road networks of Chengdu and Shanghai.}
    \label{fig:rn}
\end{figure}

For data preprocessing, we initially employed a fast map-matching algorithm \cite{Yang2018FastMM} to convert GPS traces into routes on the respective road network. We then cleaned the data, eliminating routes that contained loops and those with too few links ({\em i.e.}, less than ten links). Subsequently, the refined Chengdu dataset contained 93,125 routes, while the Shanghai dataset comprised 24,468 routes. The road networks of Chengdu and Shanghai are visually represented in Figure \ref{fig:rn}, and the key network statistics are summarized in Table \ref{tab:network}.

\renewcommand{\arraystretch}{1}
\begin{table}[h] 
\centering
\caption{Statistics of Chengdu and Shanghai road networks. (M)ID: (Mean) in-degree, (M)OD: (Mean) out-degree.}
\resizebox{0.5\textwidth}{!}{
\begin{tabular}{lcccccc}
\toprule
         & Nodes & Edges & MID / MOD & Max ID/OD, Min ID/OD  & Density \\ \midrule
Chengdu  & 2832  & 6506  & 2.297     & 4, 0                        & 8.11e-4 \\ 
Shanghai & 320   & 714   & 2.231     & 4, 1                          & 6.99e-3 \\ \bottomrule
\end{tabular}
}
\label{tab:network}
\end{table}

\vspace{-8pt}
\subsection{Baseline Methods}

In this study, we compare our approach with several established baselines to evaluate model performance:

\begin{itemize}[leftmargin=*,noitemsep]
    \item Markov: The Markov model is a widely-used sequential prediction method. It bases its route forecasting on observed transition patterns between road links.
    \item Dijkstra \cite{dijkstra1959note}: Dijkstra's algorithm is a prominent method for finding the shortest path in a graph, where the path length is assumed to be the sum of link lengths. This baseline can only work when the exact goal location is given.
    \item RNN \cite{rumelhart1986learning}: The RNN is an artificial neural network that recognizes patterns in sequential data. It accomplishes this by utilizing internal memory to process arbitrary sequences of inputs, making it effective for predicting future routes.
    \item GRU \cite{cho2014learning}: The GRU is a type of RNN that utilizes gating mechanisms to capture long-term dependencies in the data, thereby improving the model performance for trajectory or route prediction.
    \item LSTM \cite{hochreiter1997long}: As another variant of RNNs, the LSTM learns long-term dependencies in data by implementing a special architecture consisting of a series of memory cells, which effectively control the flow of information.
    \item NetTraj \cite{liang2021nettraj}: NetTraj is an advanced network-based trajectory prediction model specifically designed for predicting future movements in road networks. By integrating the Graph Attention Network (GAT) with LSTM, it leverages the spatial structure of road networks and historical trajectory data for robust future trajectory predictions.
    \item RCM-BC \cite{zhao2023deep}: RCM-BC is a behavioral cloning approach designed for route choice modeling in sequential decision-making scenarios. It employs supervised learning to create a policy that maps states to actions based on observed behavior to predict future routes. This baseline also requires knowledge of the exact goal location.
    \item RoPT \cite{xiong2025ropt}: RoPT combines stacked GCN layers with a Transformer encoder to capture spatial and long-range temporal dependencies.
\end{itemize}

\renewcommand{\arraystretch}{1.1}
\begin{table*}[t]
\centering
\caption{Performance comparison of different methods.}
\resizebox{\textwidth}{!}{
\Large
\begin{tabular}{l|ccccccccc|ccccccccc}
\toprule 
Main Results & \multicolumn{9}{c|}{Chengdu} & \multicolumn{9}{c}{Shanghai} \\ \midrule
\multirow{2}{*}{\begin{tabular}[c]{@{}c@{}}NoGoal\end{tabular}} & \multicolumn{3}{c|}{Link-level} & \multicolumn{6}{c|}{Route-level} & \multicolumn{3}{c|}{Link-level} & \multicolumn{6}{c}{Route-level} \\ 
 & R@1 & R@5 & \multicolumn{1}{c|}{R@10} & R@1 & R@5 & R@10 & M@1 & M@5 & M@10 & R@1 & R@5 & \multicolumn{1}{c|}{R@10} & R@1 & R@5 & R@10 & M@1 & M@5 & M@10 \\ \midrule
Markov & 0.696 & 0.698 & \multicolumn{1}{c|}{0.699} & 0.466 & 0.468 & 0.468 & 0.466 & 0.466 & 0.467 & 0.633 & 0.634 & \multicolumn{1}{c|}{0.635} & 0.448 & 0.448 & 0.449 & 0.448 & 0.448 & 0.448 \\
RNN & 0.812 & 0.878 & \multicolumn{1}{c|}{0.914} & 0.665 & 0.840 & 0.888 & 0.665 & 0.733 & 0.739 & 0.709 & 0.789 & \multicolumn{1}{c|}{0.837} & 0.542 & 0.719 & 0.783 & 0.542 & 0.605 & 0.613 \\
GRU & 0.799 & 0.864 & \multicolumn{1}{c|}{0.902} & 0.650 & 0.825 & 0.872 & 0.650 & 0.718 & 0.724 & 0.709 & 0.790 & \multicolumn{1}{c|}{0.839} & 0.544 & 0.718 & 0.785 & 0.544 & 0.605 & 0.614 \\
LSTM & 0.803 & 0.868 & \multicolumn{1}{c|}{0.905} & 0.656 & 0.828 & 0.877 & 0.656 & 0.723 & 0.729 & 0.700 & 0.779 & \multicolumn{1}{c|}{0.831} & 0.537 & 0.706 & 0.775 & 0.537 & 0.596 & 0.605 \\
NetTraj & 0.809 & 0.874 & \multicolumn{1}{c|}{0.909} & 0.662 & 0.836 & 0.882 & 0.662 & 0.730 & 0.735 & 0.709 & 0.788 & \multicolumn{1}{c|}{0.836} & 0.547 & 0.717 & 0.781 & 0.547 & 0.606 & 0.615 \\ 
RoPT                    & 0.824          & 0.894          & \multicolumn{1}{c|}{0.929}          & 0.675          & 0.850          & 0.895          & 0.675          & 0.747          & 0.750          & 0.715          & 0.800          & \multicolumn{1}{c|}{0.848}          & 0.560          & 0.715          & 0.770          & 0.560          & 0.615          & 0.622          \\
\textbf{RouteKG} & \textbf{0.841} & \textbf{0.940} & \multicolumn{1}{c|}{\textbf{0.968}} & \textbf{0.696} & \textbf{0.885} & \textbf{0.931} & \textbf{0.696} & \textbf{0.762} & \textbf{0.768} & \textbf{0.724} & \textbf{0.865} & \multicolumn{1}{c|}{\textbf{0.909}} & \textbf{0.563} & \textbf{0.762} & \textbf{0.831} & \textbf{0.563} & \textbf{0.624} & \textbf{0.634} \\ \midrule
\multirow{2}{*}{GoalD} & \multicolumn{3}{c|}{Link-level} & \multicolumn{6}{c|}{Route-level} & \multicolumn{3}{c|}{Link-level} & \multicolumn{6}{c}{Route-level} \\ 
 & R@1 & R@5 & \multicolumn{1}{c|}{R@10} & R@1 & R@5 & R@10 & M@1 & M@5 & M@10 & R@1 & R@5 & \multicolumn{1}{c|}{R@10} & R@1 & R@5 & R@10 & M@1 & M@5 & M@10 \\ \midrule
RNN & 0.853 & 0.911 & \multicolumn{1}{c|}{0.939} & 0.718 & 0.881 & 0.918 & 0.718 & 0.783 & 0.787 & 0.777 & 0.850 & \multicolumn{1}{c|}{0.893} & 0.621 & 0.788 & 0.849 & 0.621 & 0.683 & 0.691 \\
GRU & 0.843 & 0.902 & \multicolumn{1}{c|}{0.931} & 0.708 & 0.868 & 0.908 & 0.708 & 0.772 & 0.776 & 0.780 & 0.852 & \multicolumn{1}{c|}{0.890} & 0.625 & 0.791 & 0.844 & 0.625 & 0.687 & 0.693 \\
LSTM & 0.852 & 0.912 & \multicolumn{1}{c|}{0.936} & 0.727 & 0.882 & 0.914 & 0.727 & 0.788 & 0.792 & 0.794 & 0.859 & \multicolumn{1}{c|}{0.893} & 0.650 & 0.801 & 0.848 & 0.650 & 0.706 & 0.712 \\
NetTraj & 0.868 & 0.923 & \multicolumn{1}{c|}{0.949} & 0.741 & 0.896 & 0.929 & 0.741 & 0.802 & 0.806 & 0.803 & 0.871 & \multicolumn{1}{c|}{0.906} & 0.656 & 0.816 & 0.865 & 0.656 & 0.715 & 0.722 \\ 
RoPT                    & 0.880          & 0.935          & \multicolumn{1}{c|}{0.960}          & 0.755          & 0.910          & 0.942          & 0.755          & 0.810          & 0.815          & 0.815          & 0.882          & \multicolumn{1}{c|}{0.923}          & 0.668          & 0.828          & 0.880          & 0.668          & 0.725          & 0.728          \\
\textbf{RouteKG} & \textbf{0.916} & \textbf{0.978} & \multicolumn{1}{c|}{\textbf{0.988}} & \textbf{0.815} & \textbf{0.953} & \textbf{0.974} & \textbf{0.815} & \textbf{0.866} & \textbf{0.869} & \textbf{0.843} & \textbf{0.946} & \multicolumn{1}{c|}{\textbf{0.963}} & \textbf{0.723} & \textbf{0.894} & \textbf{0.918} & \textbf{0.723} & \textbf{0.780} & \textbf{0.784} \\ \midrule 
\multirow{2}{*}{Goal} & \multicolumn{3}{c|}{Link-level} & \multicolumn{6}{c|}{Route-level} & \multicolumn{3}{c|}{Link-level} & \multicolumn{6}{c}{Route-level} \\ 
 & R@1 & R@5 & \multicolumn{1}{c|}{R@10} & R@1 & R@5 & R@10 & M@1 & M@5 & M@10 & R@1 & R@5 & \multicolumn{1}{c|}{R@10} & R@1 & R@5 & R@10 & M@1 & M@5 & M@10 \\ \midrule
Dijkstra & 0.737 & -- & \multicolumn{1}{c|}{--} & 0.715 & -- & -- & -- & -- & -- & 0.724 & -- & \multicolumn{1}{c|}{--} & 0.703 & -- & -- & -- & -- & -- \\
RNN & 0.866 & 0.916 & \multicolumn{1}{c|}{0.941} & 0.755 & 0.892 & 0.924 & 0.755 & 0.808 & 0.812 & 0.862 & 0.902 & \multicolumn{1}{c|}{0.926} & 0.761 & 0.861 & 0.892 & 0.761 & 0.799 & 0.803 \\
GRU & 0.858 & 0.912 & \multicolumn{1}{c|}{0.939} & 0.736 & 0.883 & 0.919 & 0.736 & 0.794 & 0.798 & 0.859 & 0.900 & \multicolumn{1}{c|}{0.921} & 0.755 & 0.861 & 0.887 & 0.755 & 0.796 & 0.799 \\
LSTM & 0.872 & 0.912 & \multicolumn{1}{c|}{0.938} & 0.782 & 0.888 & 0.920 & 0.782 & 0.823 & 0.826 & 0.878 & 0.908 & \multicolumn{1}{c|}{0.929} & 0.804 & 0.877 & 0.904 & 0.804 & 0.830 & 0.833 \\
NetTraj & 0.876 & 0.918 & \multicolumn{1}{c|}{0.941} & 0.782 & 0.894 & 0.923 & 0.782 & 0.825 & 0.829 & 0.883 & 0.919 & \multicolumn{1}{c|}{0.938} & 0.790 & 0.884 & 0.910 & 0.790 & 0.826 & 0.829 \\
RCM-BC & 0.784 & 0.933 & \multicolumn{1}{c|}{0.956} & 0.669 & 0.880 & 0.918 & 0.669 & 0.754 & 0.760 & 0.827 & 0.936 & \multicolumn{1}{c|}{0.954} & 0.748 & 0.908 & 0.934 & 0.748 & 0.817 & 0.820 \\ 
RoPT                    & 0.890          & 0.950          & \multicolumn{1}{c|}{0.970}          & 0.795          & 0.907          & 0.936          & 0.795          & 0.834          & 0.838          & 0.895          & 0.932          & \multicolumn{1}{c|}{0.950}          & 0.803          & 0.897          & 0.922          & 0.803          & 0.838          & 0.842          \\
\textbf{RouteKG} & \textbf{0.974} & \textbf{0.991} & \multicolumn{1}{c|}{\textbf{0.995}} & \textbf{0.958} & \textbf{0.983} & \textbf{0.988} & \textbf{0.958} & \textbf{0.967} & \textbf{0.968} & \textbf{0.945} & \textbf{0.979} & \multicolumn{1}{c|}{\textbf{0.984}} & \textbf{0.915} & \textbf{0.959} & \textbf{0.969} & \textbf{0.915} & \textbf{0.932} & \textbf{0.933} \\ \bottomrule 
\end{tabular}
}
\label{tab:mainresults}
\vspace{-3mm}
\end{table*}

\subsection{Main Results} \label{sec:results:main}
\subsubsection{Experimental Settings}

To comprehensively assess model performance, we design experiments based on the three subproblems as defined in Section~\ref{subsec:pre:problem}: (1) route prediction with unknown goal $\mathcal{F}_1$, (2) route prediction with goal direction only $\mathcal{F}_2$, and (3) route prediction with complete goal information $\mathcal{F}_3$. We refer to these three subproblems as \textit{NoGoal}, \textit{GoalD}, and \textit{Goal}. They reflect varying degrees of information availability regarding the road user's intended destination, and represent a broad range of real-world application scenarios. For instance, a system might not know the user's exact destination due to privacy concerns, but could have access to more general information, such as the goal direction. 

Most of the baseline models are designed for the \textit{NoGoal} scenario, but two of them (Dijkstra and RCM-BC) are for the \textit{Goal} scenario only. Unlike these baselines, RouteKG requires the goal direction information. Therefore, specific model implementations are needed to incorporate the available goal information into different models under different scenarios. Under the \textit{NoGoal} scenario, the goal direction is unknown, but we can still estimate it based on the observed route. Consequently, the estimated goal direction is used in RouteKG under \textit{NoGoal}. Under GoalD, the actual goal direction is used instead of the estimated one in RouteKG. For other deep learning baseline models (except for RCM-BC), we concatenate the embedding of goal directions with the respective model's inputs. Similarly, under the \textit{Goal} scenario, the same concatenation strategy can be used for the baseline models, enriching them with complete goal information. In RouteKG, we add the embedding of the goal location directly to the embedding of the last link in the observed route. 
The ConsistentWith relation in the KG is constructed using observed trajectories to compute link co-occurrence. This requires access to full observed trajectories $\mathcal{X}^o$ during training, but not at inference time. For the \textit{EGoalD} scenario, both training and inference require the observed route prefix $\mathcal{X}^o$ to estimate the goal direction. In contrast, under the \textit{GoalD} and \textit{Goal} settings, both training and inference rely only on the last observed link $e^\Gamma$ to predict the future route. This makes RouteKG applicable even in cases with limited route history.

In our main experiments, the input observed route length is set as $\Gamma=10$ and the output future route length as $\Gamma^\prime=5$. For model evaluation, the datasets are partitioned into training, validation, and test subsets in a 6:2:2 ratio. Different models are evaluated under the \textit{NoGoal}, \textit{GoalD}, and \textit{Goal} scenarios, using both the ``link-level'' and ``route-level'' metrics. Link-level assessment has practical implications, particularly for tasks related to traffic flows, while route-level evaluation offers valuable information for routing applications. Specifically, we utilize Recall and Mean Reciprocal Rank (MRR), two prevalent metrics. Recall measures the ratio of relevant items retrieved from all relevant items, indicating the system's capacity to fetch desired information. MRR, on the other hand, evaluates the rank position of the correct answer, computing the average reciprocal rank of the highest-ranked correct answer across queries. A higher MRR signifies superior performance. These metrics provide insights into model effectiveness and ranking quality and are useful tools for assessing and enhancing system performance.

\begin{figure*}[h]
    \centering
    \includegraphics[width=.75\textwidth]{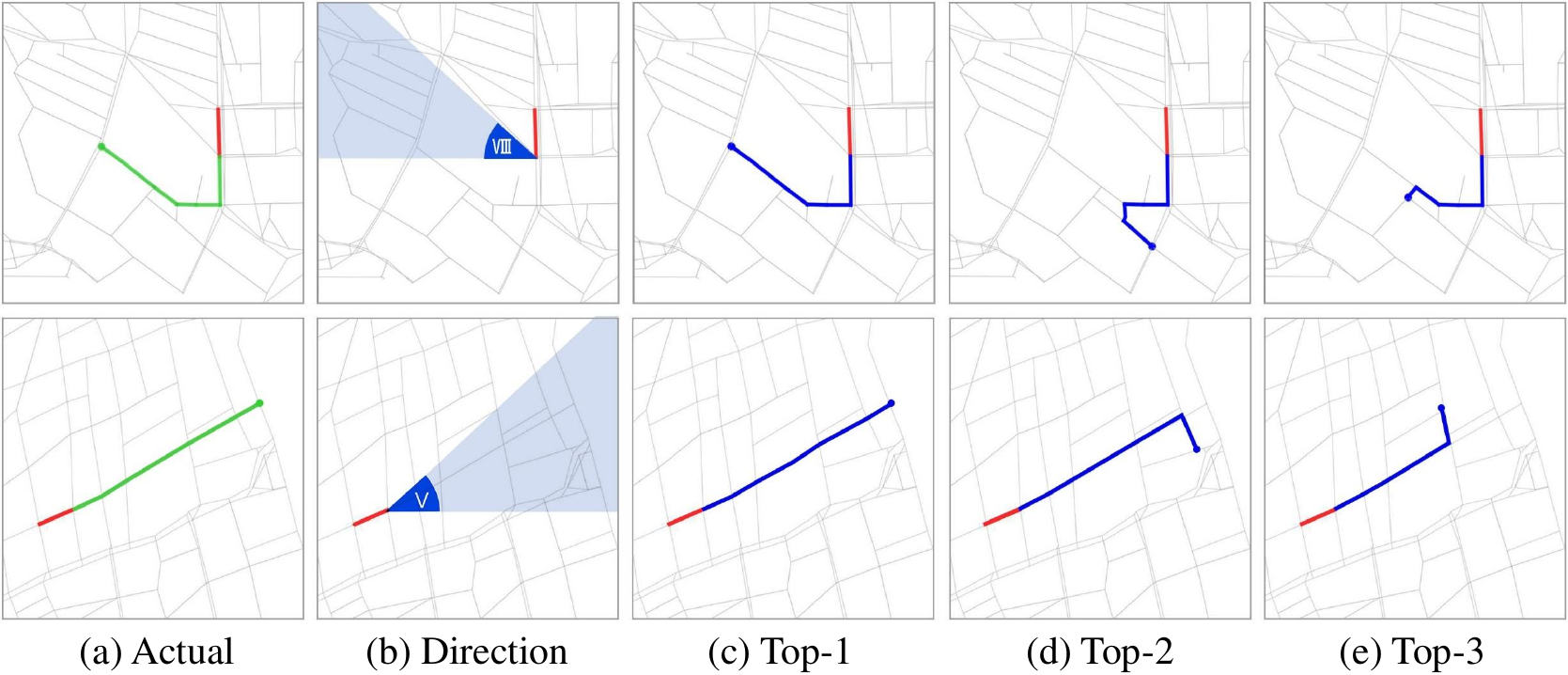}
    \caption{Example results in Chengdu (upper) and Shanghai (lower), with the red line indicating the last observed link, the green line the actual future route, and the blue lines are the predicted future routes.}
    \label{fig:qual}
    \vspace{-3mm}
\end{figure*}

To define evaluation metrics, we consider the top-$k$ predictions for the $i$-th observed route $\{\widehat{x^f_{i, k}}\}_{k=1}^K$=$\{\{\widehat{e_{i, k}^j}\}_{j=\Gamma+1}^{\Gamma+\Gamma^\prime}\}_{k=1}^K$ and the actual $i$-th future route $x_i^f=\{e_i^j\}_{j=\Gamma+1}^{\Gamma+\Gamma^\prime}$. 
The link-level recall R@K is defined as
$R@K=\frac{1}{|\mathcal{X}|}\sum_{i=1}^{|\mathcal{X}|}\max_{k=1}^{K}\left[\frac{1}{\Gamma^\prime}\sum_{j=\Gamma+1}^{\Gamma+\Gamma^\prime}\mathbb{I}(\widehat{e_{i, k}^j}=e_i^j)\right].$
where $\mathbb{I}(a=b)=\begin{cases} 1 & \text{if } a = b \\ 0 & \text{otherwise} \end{cases}$ is the indicator function.

Similarly, the route-level recall R@K is defined as
$R@K=\frac{1}{|\mathcal{X}|}\sum_{i=1}^{|\mathcal{X}|}\max_{k=1}^{K}\left[\mathbb{I}(\sum_{j=\Gamma+1}^{\Gamma+\Gamma^\prime}\mathbb{I}(\widehat{e_{i, k}^j}=e_i^j), \Gamma^\prime)\right].$

We also compute the route-level MRR of the top-$k$ predictions, M@K, as follows:
$M@K=\frac{1}{|\mathcal{X}|}\sum_{i=1}^{|\mathcal{X}|}\sum_{k=1}^{K}\frac{1}{k}\left[\mathbb{I}(\sum_{j=\Gamma+1}^{\Gamma+\Gamma^\prime}\mathbb{I}(\widehat{e_{i, k}^j}=e_i^j), \Gamma^\prime )\right].$

The experiments are conducted on a Ubuntu server with the Python 3.6 environment. The deep learning computations are performed using the PyTorch framework. The server's hardware specifications include an Intel(R) Xeon(R) Platinum 8375C CPU with a clock speed of 2.90GHz, coupled with 8 NVIDIA GeForce RTX 3090 GPUs, each featuring 24GB of memory. To ensure the robustness and generalizability of our model, hyperparameters are tuned based on the performance of the validation set, which is crucial for balancing the bias-variance trade-off and optimizing the model's performance. All hyperparameter settings are detailed in Appendix~\ref{apx:hyperparams}.

\subsubsection{Main Results Analysis}

Table \ref{tab:mainresults} shows a comparison of the accuracy of the different methods in predicting future routes on two real-world datasets under three scenarios. 
Overall, RouteKG consistently outperforms all baselines across all evaluation metrics.
It is observed that for all models, the route-level prediction accuracy is lower than the link-level prediction accuracy. This underscores the importance of modeling the consistency between different road links. Comparing the different models, we find that deep learning methods achieve higher accuracy in general and can be further enhanced by integrating additional information. For instance, models like NetTraj and RouteKG, which incorporate spatial data, outperform simpler models like RNN and its variants. Remarkably, RouteKG outperforms the NetTraj model and other baselines, even without extra information, which highlights the effectiveness of our approach to integrate KG for future route prediction. Comparing under different experimental settings, intuitively, introducing more goal information progressively improves overall accuracy. In particular, RouteKG's prediction accuracy is greatly improved after the gradual incorporation of goal information, and at the same time, it also has an accuracy improvement of about 5.41\% in comparison with the optimal baseline under the NoGoal condition, which proves the effectiveness of RouteKG in utilizing goal information and its applicability under various conditions. Lastly, comparing different datasets, the prediction accuracy of the Chengdu dataset is better than that of the Shanghai dataset, likely because of the larger data size of the former.

To provide an intuitive understanding of the prediction results, we show two qualitative examples of RouteKG outputs under \textit{NoGoal} in Figure~\ref{fig:qual}. Each contains the last observed link alongside the estimated direction and top-3 predictions. Although certain future routes may display peculiar turns due to constraints imposed by the road network, most predicted future routes exhibit a correct heading based on the estimated goal direction. An important observation from the first example is the misalignment between the links adjacent to the last observed link and the predicted direction. Consequently, the predicted road links are initially constrained on the road network. However, as the prediction progresses, subsequent steps are adjusted to align with the route's predicted direction.

\begin{figure}[h]
    \centering
    \includegraphics[width=.48\textwidth]{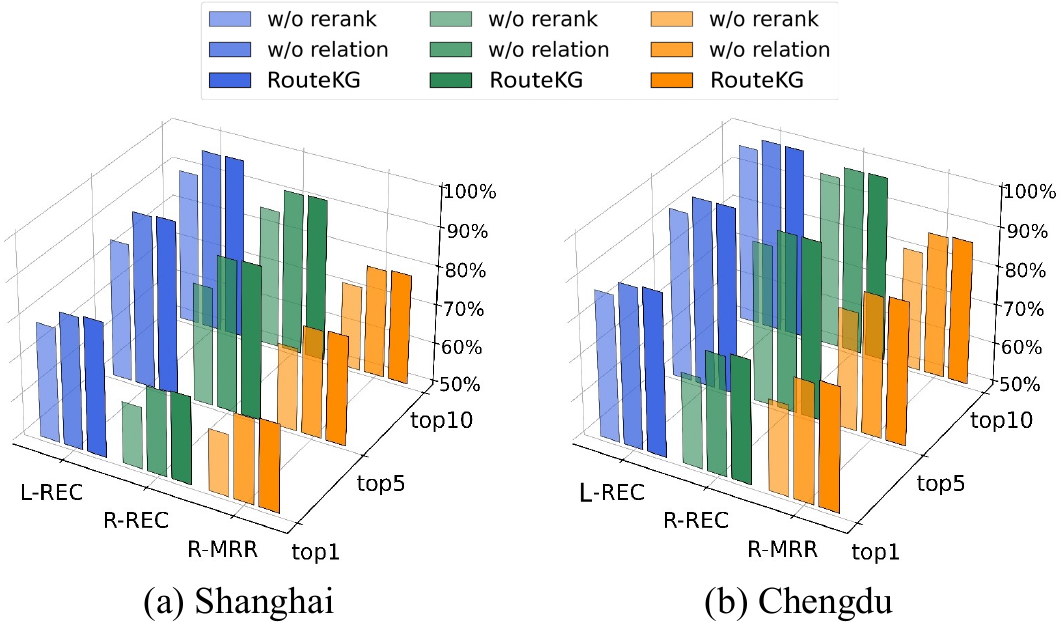}
    \caption{Ablation analysis results with goal direction.}
    \label{fig:abl_goald}
    \vspace{-3mm}
\end{figure}

\subsection{Ablation Analysis}

This section analyzes the results of the ablation experiments. Specifically, we focus on the analysis performed on RouteKG under \textit{GoalD}. By conducting these ablation experiments, we can gain insight into the importance of each component of the model and its contribution to the overall predictive capabilities.

Figure~\ref{fig:abl_goald} compares the performance of RouteKG with its two ablation variants, where L-Rec denotes link-level recall, R-Rec denotes route-level recall, and R-MRR stands for route-level mean reciprocal rank. Notably, RouteKG w/o rerank removes the \textit{Rank Refinement Module}. Experimental results show that removing this module significantly reduces prediction performance. This suggests the interconnected nature of link choices, emphasizing the need for a module to model route consistency and choice correlation. This highlights the module's indispensability. Remarkably, even without reranking, RouteKG still outperforms most benchmark methods, particularly in the top 5 and top 10 predictions. This demonstrates RouteKG's efficacy in identifying potential future routes, reinforcing the importance of integrating the ranking refinement module for enhancing top-1 predictions. The detailed ablation analysis results are shown in Appendix~\ref{apx:abl}.

RouteKG w/o relation denotes the RouteKG model removes the KG representation learning. The observed performance drop in this variant is less pronounced compared to the removal of the \textit{Rank Refinement Module}. This indicates that although KG representation learning is beneficial to the route prediction process, it acts more as an auxiliary component. The substantial effectiveness of using KGC alone in predicting future routes underscores the suitability of approaching future route prediction as a KGC problem.

\begin{table}[h]
\centering
\caption{Ablation on relation types (NoGoal, Chengdu, link-level R@1).}
\resizebox{.65\linewidth}{!}{
\begin{tabular}{lcc}
\toprule
Setting & R@1 & $\Delta$ \\
\midrule
All relations          & \textbf{0.841} & -- \\
\quad w/o DirectionTo  & 0.820 & $-2.1\%$ \\
\quad w/o ConsistentWith & 0.824 & $-1.7\%$ \\
\quad w/o ConnectBy    & 0.831 & $-1.0\%$ \\
\quad w/o DistanceTo   & 0.832 & $-0.9\%$ \\
\bottomrule
\end{tabular}
}
\label{tab:rel_ablation}
\end{table}

We further conduct ablation experiments on individual relation types to understand their respective contributions. As shown in Table~\ref{tab:rel_ablation}, removing any single relation leads to a modest drop in performance (within 2\% in link-level R@1), with \textit{DirectionTo} being the most influential. These results suggest that while the model can learn effectively through the KGC formulation, KG representation learning still provides complementary relational priors that enhance prediction, especially under sparse supervision scenarios.

\subsection{Sensitivity Analysis}

\begin{figure}[h]
    \centering
    \includegraphics[width=.48\textwidth]{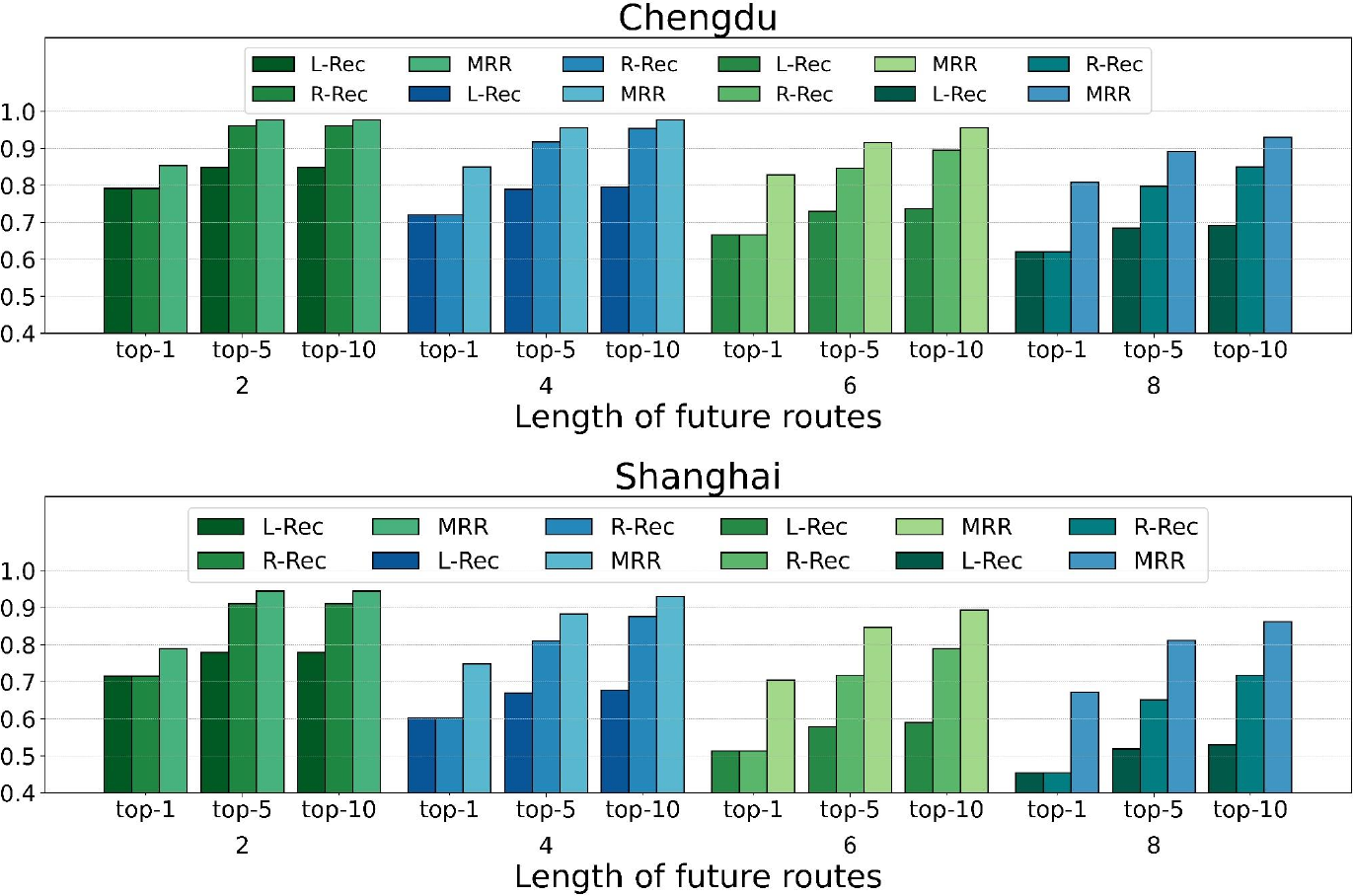}
    \caption{$\Gamma^\prime$ sensitivities on Shanghai under \textit{NoGoal}.}
    \label{fig:shed_prelen}
\end{figure}

This section presents a sensitivity analysis to assess the robustness and reliability of our model under various parameter settings. We first investigate the model performance under varying lengths of future routes to be predicted, denoted as $\Gamma^\prime$. As depicted in Figure~\ref{fig:shed_prelen}, by altering $\Gamma^\prime$ from 2 to 8, there is a noticeable trend of declining performance with increasing $\Gamma^\prime$ values. This indicates that predicting longer routes becomes progressively challenging due to an expanded candidate space and uncertainty, particularly when lacking goal information.

\renewcommand{\arraystretch}{1.1}
\begin{table}[h]
\centering
\caption{Performance of $\tau$-Pruned and 1-Greedy decoding strategies in Chengdu. Both maintain high accuracy with reduced computational cost.}
\resizebox{\linewidth}{!}{
\begin{tabular}{lccccccccc}
\toprule
\multirow{2}{*}{} & \multicolumn{3}{c}{Link-level}            & \multicolumn{6}{c}{Route-level}               \\
                  & R@1                       & R@5   & R@10  & R@1   & R@5   & R@10  & M@1   & M@5   & M@10  \\ \midrule
NoGoal            & 0.841                     & 0.940 & 0.968 & 0.696 & 0.885 & 0.931 & 0.696 & 0.762 & 0.768 \\
$\tau$-Pruned     & 0.830                     & 0.935 & 0.965 & 0.690 & 0.880 & 0.925 & 0.690 & 0.758 & 0.765 \\
1-Greedy          & \multicolumn{1}{l}{0.821} & $-$  & $-$  & 0.674 & $-$  & $-$  & 0.674 & $-$  & $-$  \\
NetTraj           & 0.809                     & 0.874 & 0.909 & 0.662 & 0.836 & 0.882 & 0.662 & 0.730 & 0.735 \\ \midrule
GoalD             & 0.916                     & 0.978 & 0.988 & 0.815 & 0.953 & 0.974 & 0.815 & 0.866 & 0.869 \\
$\tau$-Pruned     & 0.905                     & 0.970 & 0.987 & 0.810 & 0.950 & 0.972 & 0.810 & 0.860 & 0.868 \\
1-Greedy          & \multicolumn{1}{l}{0.898} & $-$  & $-$  & 0.794 & $-$  & $-$  & 0.794 & $-$  & $-$  \\
NetTraj           & 0.868                     & 0.923 & 0.949 & 0.741 & 0.896 & 0.929 & 0.741 & 0.802 & 0.806 \\ \midrule
Goal              & 0.974                     & 0.991 & 0.995 & 0.958 & 0.983 & 0.988 & 0.958 & 0.967 & 0.968 \\
$\tau$-Pruned     & 0.965                     & 0.990 & 0.994 & 0.955 & 0.980 & 0.986 & 0.955 & 0.965 & 0.966 \\
1-Greedy          & \multicolumn{1}{l}{0.952} & $-$  & $-$  & 0.948 & $-$  & $-$  & 0.948 & $-$  & $-$  \\
NetTraj           & 0.876                     & 0.918 & 0.941 & 0.782 & 0.894 & 0.923 & 0.782 & 0.825 & 0.829 \\ \bottomrule
\end{tabular}
}
\label{tab:pruned}
\end{table}

We further evaluate two heuristic decoding strategies to understand their trade-offs in accuracy and computational efficiency: (1) $\tau$-Pruned retains only the next-step candidates whose transition probability exceeds a pruning threshold $\tau$ (set as 0.2). (2) 1-Greedy is a fully greedy decoding strategy that selects only the most probable next-step link at each prediction step, effectively reducing the candidate set size to one per step.

Table~\ref{tab:pruned} summarizes the performance of these variants on the Chengdu dataset. While both heuristics lead to slight reductions in performance compared to the full search, they still maintain competitive accuracy. For example, $\tau$-Pruned achieves a route-level R@10 of 0.986 (vs. 0.988 in full search), and 1-Greedy reaches 0.948 with goal information. These results highlight that RouteKG retains strong predictive performance even with aggressive pruning, validating the model's robustness and generalization under restricted search space.

\subsection{Efficiency Analysis}

\begin{figure}[ht]
    \centering
    \includegraphics[width=.48\textwidth]{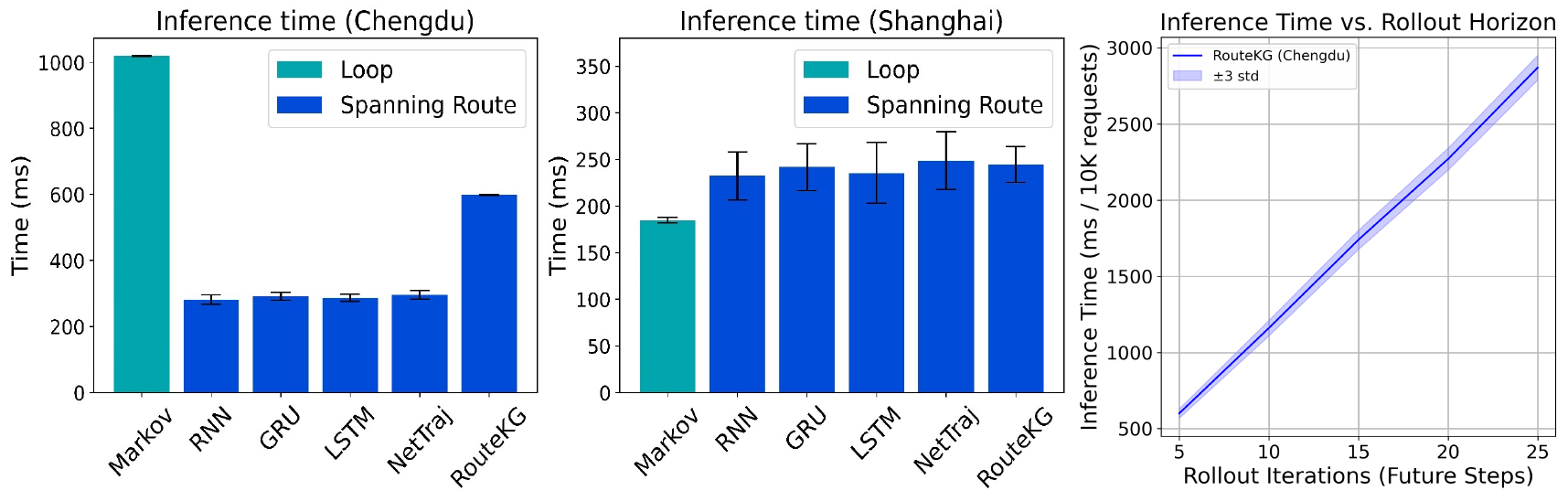}
    \caption{Left two: Inference time per 10K requests for different methods in Chengdu and Shanghai (mean ± std). Right: Inference time vs. rollout horizons, iterative top-1 rollouts show linear runtime growth.}
    \label{fig:inftime}
\end{figure}

Efficient route prediction in transportation systems is crucial for ensuring prompt responses for system operators and road users. To assess the model efficiency, we analyze the inference time of various models in two datasets. Figure \ref{fig:inftime} delineates the inference times across various models. Note that the results from Dijkstra and RCM-BC are omitted due to their overly long inference times. All baselines utilize the \textit{Spanning Route} algorithm to generate future routes except the Markov model, which uses pre-computed transition probabilities to sample and generate the top-$k$ predictions through $k$ iterations.

\renewcommand{\arraystretch}{1}
\begin{table*}[t]
\centering
\caption{Traffic flow estimation results on Chengdu and Shanghai datasets. (mean ± std)}
\resizebox{.85\textwidth}{!}{
\begin{tabular}{l|ccc|ccc}
\toprule
Traffic Flow & \multicolumn{3}{c|}{Chengdu} & \multicolumn{3}{c}{Shanghai} \\ \midrule
NoGoal & MAE & RMSE & $\mathrm{R}^2$ & MAE & RMSE & $\mathrm{R}^2$ \\ \midrule 
Markov & 7.849 ± 0.003 & 26.104 ± 0.010 & 0.820 ± 0.000 & 20.758 ± 0.012 & 42.040 ± 0.021 & 0.716 ± 0.000 \\

RNN & 3.774 ± 0.016 & 12.018 ± 0.109 & 0.962 ± 0.001 & 12.062 ± 0.167 & 22.711 ± 0.183 & 0.917 ± 0.001 \\
GRU & 3.974 ± 0.014 & 12.180 ± 0.115 & 0.961 ± 0.001 & 12.453 ± 0.153 & 23.674 ± 0.207 & 0.910 ± 0.002 \\
LSTM & 3.961 ± 0.008 & 12.517 ± 0.100 & 0.959 ± 0.001 & 12.705 ± 0.115 & 23.980 ± 0.137 & 0.908 ± 0.001 \\
NetTraj & 3.777 ± 0.014 & 11.896 ± 0.121 & 0.963 ± 0.001 & 12.333 ± 0.137 & 23.030 ± 0.201 & 0.915 ± 0.001 \\ 
RoPT & 3.15 ± 0.02 & 9.52 ± 0.11 & 0.976 ± 0.001 & 10.98 ± 0.18 & 20.90 ± 0.30 & 0.932 ± 0.002 \\ 
\textbf{RouteKG} & \textbf{2.464 ± 0.030} & \textbf{6.725 ± 0.116} & \textbf{0.988 ± 0.000} & \textbf{8.178 ± 0.200} & \textbf{15.330 ± 0.537} & \textbf{0.962 ± 0.003} \\
\midrule 
GoalD & MAE & RMSE & $\mathrm{R}^2$ & MAE & RMSE & $\mathrm{R}^2$ \\ \midrule 
RNN & 3.121 ± 0.010 & 11.034 ± 0.114 & 0.968 ± 0.001 & 9.520 ± 0.136 & 17.287 ± 0.182 & 0.952 ± 0.001 \\
GRU & 3.226 ± 0.013 & 11.051 ± 0.120 & 0.968 ± 0.001 & 9.108 ± 0.153 & 16.400 ± 0.219 & 0.957 ± 0.001 \\
LSTM & 3.111 ± 0.009 & 10.995 ± 0.110 & 0.968 ± 0.001 & 8.390 ± 0.167 & 14.731 ± 0.277 & 0.965 ± 0.001 \\
NetTraj & 2.948 ± 0.003 & 10.796 ± 0.130 & 0.969 ± 0.001 & 8.006 ± 0.176 & 14.311 ± 0.307 & 0.967 ± 0.001 \\  
RoPT & 2.35 ± 0.02 & 8.27 ± 0.12 & 0.982 ± 0.001 & 6.22 ± 0.12 & 10.45 ± 0.21 & 0.981 ± 0.001 \\ 
\textbf{RouteKG} & \textbf{1.688 ± 0.032} & \textbf{6.237 ± 0.161} & \textbf{0.990 ± 0.001} & \textbf{4.682 ± 0.091} & \textbf{7.237 ± 0.284} & \textbf{0.992 ± 0.001} \\ \midrule
Goal & MAE & RMSE & $\mathrm{R}^2$ & MAE & RMSE & $\mathrm{R}^2$ \\ \midrule 
Dijkstra & 4.386 & 17.146 & 0.922 & 12.655 & 29.711 & 0.858 \\
RNN & 2.988 ± 0.007 & 10.084 ± 0.170 & 0.973 ± 0.001 & 7.200 ± 0.144 & 14.232 ± 0.248 & 0.967 ± 0.001 \\
GRU & 3.055 ± 0.011 & 10.161 ± 0.169 & 0.973 ± 0.001 & 7.100 ± 0.127 & 13.496 ± 0.237 & 0.971 ± 0.001 \\
LSTM & 2.962 ± 0.014 & 10.167 ± 0.156 & 0.973 ± 0.001 & 6.903 ± 0.117 & 13.703 ± 0.187 & 0.970 ± 0.001 \\
NetTraj & 2.899 ± 0.004 & 9.970 ± 0.175 & 0.974 ± 0.001 & 6.669 ± 0.126 & 13.604 ± 0.231 & 0.970 ± 0.001 \\
RCM-BC & 3.299 ± 0.036 & 7.923 ± 0.037 & 0.980 ± 0.000 & 4.993 ± 0.136 & 8.701 ± 0.049 & 0.988 ± 0.000 \\ 
RoPT & 2.48 ± 0.02 & 7.28 ± 0.10 & 0.985 ± 0.001 & 4.96 ± 0.10 & 9.34 ± 0.17 & 0.987 ± 0.001 \\ 
\textbf{RouteKG} & \textbf{1.012 ± 0.016} & \textbf{3.168 ± 0.096} & \textbf{0.997 ± 0.000} & \textbf{3.604 ± 0.088} & \textbf{6.340 ± 0.151} & \textbf{0.994 ± 0.000} \\ \bottomrule 
\end{tabular}
}
\label{tab:flow}
\vspace{-3mm}
\end{table*}

RouteKG demonstrates remarkable efficiency, achieving average inference times of 598.01ms and 244.47ms for every 10k requests on the Chengdu and Shanghai datasets, respectively, with standard deviations of 1.21ms and 19.35ms. In contrast, the Dijkstra model, based on dynamic programming, takes over 38s, and the RCM-BC model exceeds 1000s, rendering them impractical for real-time systems. Models utilizing the \textit{Spanning Route} algorithm (e.g., RNN, GRU, LSTM, NetTraj, RouteKG) show superior inference times, with less than 400ms in Chengdu and 250ms in Shanghai per 10k requests. RouteKG exhibits a marginally higher inference time, likely attributable to the reranking process.

Note that our framework is designed for short-term route prediction, where local transition patterns remain stable. For longer prediction horizons, directly increasing the number of steps is impractical, as it leads to exponential growth in candidate routes under the Spanning Route framework. We discuss these limitations and possible extensions in the Discussion section.
To support longer predictions efficiently, we adopt a recursive rollout strategy. In this setting, the model predicts a short future segment (e.g., 5 steps) and recursively appends the top-1 predicted links as input for the next segment, enabling longer predictions without modifying the base architecture. This strategy leverages RouteKG’s high top-1 accuracy, making it suitable for multi-step rollout.

As shown in the right panel of Figure~\ref{fig:inftime}, inference time increases approximately linearly with the number of rollout iterations, under the same computational resources. This confirms that our framework remains computationally efficient even as the prediction horizon extends. Unlike conventional methods, where prediction cost grows exponentially due to expanding search trees, our strategy avoids full enumeration by committing to a single most likely route per segment. This makes the model suitable for time-sensitive applications where latency constraints prohibit full route expansion.

\subsection{Case Study: Traffic Flow Estimation}

In this section, to demonstrate the practical use cases of RouteKG, we conduct a case study on traffic flow estimation utilizing the model's capability to generate future routes, which can then support other traffic control or management strategies. Specifically, we adopt a sampling-based method for generating future routes to maximize the utility of the top-$k$ future route predictions. Initially, the top-$k$ predictions are converted to a probability distribution using temperature scaling \cite{guo2017calibration}. Subsequently, we sample from the predicted top-$k$ future routes for each observed trajectory based on their probability distribution. The estimated link-level traffic flows are then obtained by aggregating the number of predicted future routes. To ensure robustness, we iterate the experiments ten times and report traffic flow estimation results in a mean±std format, focusing solely on the top-10 predictions for simplicity.

The effectiveness of traffic flow estimation with RouteKG is demonstrated using three standard regression metrics: Mean Absolute Error (MAE), Root Mean Squared Error (RMSE), and the coefficient of determination ($\mathrm{R}^2$), as detailed in Table~\ref{tab:flow}. RouteKG consistently outperforms in all metrics for both datasets, aligning with our main experiment results in Section~\ref{sec:results:main}. As expected, incorporating more goal information leads to improved accuracy in traffic flow predictions.

In particular, RouteKG's performance in the \textit{NoGoal} scenario significantly surpasses the baseline for both datasets, suggesting that our method of estimating moving directions and leveraging KGC is more effective than current state-of-the-art (SOTA) modeling methods. Quantitatively, it reduces MAE, RMSE, and $\mathrm{R}^2$ by 34.7\%, 43.5\%, and 2.6\%, respectively, compared to the best baseline. Under the \textit{GoalD} scenario, performance increases notably, indicating potential for future refinement in modeling future directions. Importantly, RouteKG's enhancements in traffic flow estimation, especially when including the actual future direction, are more significant than those of the baselines. This reaffirms RouteKG's integration of direction information in the KGC problem. With actual goal information incorporated, RouteKG achieves an MAE of 1, RMSE of 3, and 99.7\% in $\mathrm{R}^2$, underlining its efficacy and promise for practical applications.

To summarize, these results suggest that RouteKG is effective in traffic flow estimation, offering accurate and rapid analysis essential for real-time traffic management.

\subsection{Discussion}

While the current framework builds on a static KG to model road network relations, it is designed with a focus on building a general and flexible solution based on the fundamental elements of road networks and could flexibly incorporate both static and dynamic contextual features. Real-world factors such as rush hours, weather, road closures, and special events can significantly affect route choices. Although these factors are not explicitly modeled in the current version, the framework naturally supports their integration. 
Contextual features can be embedded into the entity representations to reflect time-dependent behaviors or region-specific characteristics. For example, embeddings related to rush hours or nearby POIs can adjust the representation of road links to capture patterns such as congestion avoidance or attraction to popular destinations. These context-aware embeddings can be learned during KG representation learning and updated continuously as new data arrives, allowing the KG to evolve and better reflect realistic routing behavior.

Additionally, disruptions such as accidents or road closures can be handled by updating the NAE matrix, which dynamically masks affected links from the candidate space during prediction. This simple mechanism enables the model to adapt to real-time changes without retraining. RouteKG is also promising for sparse road networks, where simpler topology reduces prediction complexity. Even with fewer links, spatial relations in the KG can provide useful priors. While these extensions are not included in the current implementation, they can be incorporated into the existing framework with minimal modifications and are left for future work.

Finally, although our framework is optimized for short-term predictions, directly increasing the prediction length may cause rapid growth in the search space and inference time under the Spanning Route paradigm. We explore two heuristic strategies to mitigate this issue: pruning low-probability candidates and recursively rolling out top-1 predictions in short segments. These strategies reduce computational overhead and demonstrate promising performance in our sensitivity and efficiency studies. Nonetheless, efficiently scaling to longer prediction horizons remains an open challenge and a key direction for future improvement. In particular, enabling variable-length prediction, where the model learns to determine when to terminate based on the observed partial route (e.g., via a special ``[End]'' token), represents a promising avenue for future research.

\section{Conclusion} \label{sec:conclusion}

This research presents RouteKG, a novel knowledge graph (KG) framework for short-term route prediction on road networks. It treats route prediction as a knowledge graph completion (KGC) problem. The framework constructs a KG based on the road network to facilitate KG representation learning, which is designed to capture spatial relations that are essential for various urban routing tasks. Through KGC, the learned relations can be further utilized for future route prediction. The devised \textit{Spanning Route} algorithm allows for the efficient generation of multiple possible future routes, while a \textit{Rank Refinement Module} is integrated to further leverage learned spatial relations to rerank the initial predictions, thereby achieving more accurate route prediction results.

RouteKG is evaluated using taxi trajectory data from Chengdu and Shanghai. The evaluation considers three practical scenarios with different levels of goal information availability: \textit{NoGoal}, \textit{GoalD}, and \textit{Goal}. The experiment results show that the proposed RouteKG consistently outperforms the baseline methods based on various evaluation metrics. Additionally, the model efficiency analysis highlights that route predictions can be generated in less than 500ms per 10k requests, largely thanks to the \textit{Spanning Route} algorithm, which validates the suitability of RouteKG for real-time traffic applications. To demonstrate the applicability of RouteKG beyond routing tasks, we utilize it to estimate link-level traffic flows, achieving an $\mathrm{R}^2$ value of 0.997 under the \textit{Goal} scenario.

Future research can extend this work in several ways. First, incorporating other spatial relations ({\em e.g.}, function zones, spatial regions, etc.) with urban and road network attributes can augment the scalability and generalizability of the model. This would enable the model to provide high-performance feedback for multi-functional intelligent transportation services rapidly, adapting to different tasks promptly. Second, future work can potentially enhance the \textit{Spanning Route} algorithm by integrating an n-ary tree pruning approach, offering a solution to model complexity increases exponentially with route prediction length. The optimized algorithm is anticipated to offer superior scalability and more efficient future route generation with reduced computational resources. Lastly, future research could explore more deeply how KGs can be leveraged for broader urban applications, such as integrating diverse datasets and uncovering interrelationships between them. For instance, identifying correlations between traffic patterns and population demographics could enable urban planners to better anticipate the impact of different urban development strategies, ultimately fostering smarter, more sustainable cities and improving overall urban system efficiency.

\section*{Acknowledgment}
This research is supported by National Natural Science Foundation of China (42201502), and Seed Funding for Strategic Interdisciplinary Research Scheme at The University of Hong Kong (102010057).

\section*{Appendix}

\subsection{Minibatch version of the \textbf{Spanning Route} algorithm}
\label{apx:batchalg}

Algorithm \ref{alg:srbatch} gives the pseudocode for the minibatch \textit{Spanning Route}.
\vspace{0.5cm}

\setcounter{algocf}{2}
\begin{algorithm}[h]
\caption{\textbf{Spanning Route} (minibatch).}
\SetKwInOut{Input}{Input}\SetKwInOut{Output}{Output}
\label{alg:srbatch}

\Input{~$\Gamma^\prime$ batched probability distributions $\left\{\mathrm{Pr}(\widetilde{x^{f, \gamma}}) \in \mathbb{R}^{\mathcal{B} \times |\mathcal{E}|}\right\}_{\gamma=1}^{\Gamma^\prime}$;\\
     road network $\mathbf{G}=(\mathbf{V}, \mathbf{E})$;\\ 
     NAE matrix $\mathbf{A} \in \mathbb{R}^{|\mathbf{V}| \times N_A}$; tree's degree $n$.
    }
\Output{~Top-$K$ predicted batched future routes $\{\widetilde{x_{k}^{f}}\}_{k=1}^{K}$.}
\BlankLine

\textcolor{darkgray}{// Initialize the root node.} \\
root $\leftarrow$ \textbf{CreateNewNode}(name = ``root'', parent = NIL, end\_nodes = $v_{\Gamma}^{s} \in \mathbb{R}^{\mathcal{B}}$, preds = NIL) \\
\textcolor{darkgray}{// Recursively generate a tree of future routes in a greedy manner.} \\
\For{$\gamma=1, \dots, \Gamma^\prime$}{
    \textcolor{darkgray}{// Get leaves of the current tree.} \\
    leaves $\leftarrow$ \textbf{GetLeaves}(root) \\
    \textcolor{darkgray}{// Span for each leaf.} \\
    \For{leaf $\in$ leaves}{
        \textcolor{darkgray}{// Get the adjacent edges given the end node.} \\
        $\mathcal{N}_{end\_node}^{e} \in \mathbb{R}^{\mathcal{B} \times N_A}$ = $\mathbf{A}[leaf.end\_nodes, :]$ \\
        \textcolor{darkgray}{// Get the top-$n$ adjacent edges with highest probabilities based on $\mathrm{Pr}(\widetilde{x^{f, \gamma}_{i}})$.} \\
        $\left\{e_{k}^{\Gamma+\gamma} \in \mathbb{R}^{\mathcal{B}}\right\}_{k=1}^{n}$ = \textbf{GetTopK}($\mathrm{Pr}(\widetilde{x^{f, \gamma}})[:, \mathcal{N}_{end\_node}^{e}]$, $K=n$) \\
        \textcolor{darkgray}{// Create leaf node for top-$n$ edges and add to the tree.} \\
        \For{$k=1, \dots, n$}{
            \textcolor{darkgray}{// Create leaf node and add to the tree.} \\
            node = \textbf{CreateNewNode}(name=``$k$'', parent=leaf, end\_node=$e_{k}^{\Gamma+\gamma}[1]  \in \mathbb{R}^{\mathcal{B}} $, pred=$e_{k}^{\Gamma+\gamma} \in \mathbb{R}^{\mathcal{B}}$)
        }
    }
    
}
leaves $\leftarrow$ \textbf{GetLeaves}(root) \\
\textcolor{darkgray}{// Traverse the tree to get top-$K$ future routes.} \\
\For{$k=1, \dots, K$}{
    \textcolor{darkgray}{// Get the path from root to the $k$-th leaf.} \\
    $\text{path}_{k}$ = \textbf{Traverse}(root, leaves[$k$]) \\ 
    \textcolor{darkgray}{// Get the generated $k$-th route.} \\
    $\widetilde{x_{k}^{f}} \in \mathbb{R}^{\mathcal{B} \times \Gamma^\prime}$ = $\left\{\text{path}_{k}\text{[i].pred} \in \mathbb{R}^{\mathcal{B}}\right\}_{i=1}^{\Gamma^\prime}$ \\ 
}
\end{algorithm}

\subsection{Hyperparameters}
\label{apx:hyperparams}

This study applied consistent experimental configurations to both datasets to ensure reliable and comparable results. All hyperparameters were selected by grid search on a held‐out validation set, ensuring that our final choices lie within the searched ranges. Below, we summarize both the search grids and the selected values.

We used a batch size of 2048 and trained for up to 10,000 epochs, with early stopping after 100 epochs without validation improvement. All embedding dimensions 
\(\delta_{\mathcal{E}}, \delta_{\mathcal{R}^c}, \delta_{\mathcal{R}^s}, \delta_{\mathcal{R}^a}, \delta_{\mathcal{R}^d}\) 
were searched over \(\{32,64,128\}\) and set to 64 in the final model. The Adam optimizer was used with weight decay chosen from \(\{10^{-4},10^{-3},10^{-2}\}\); we selected \(10^{-2}\). Learning rates were searched over \(\{10^{-4},10^{-3},10^{-2}\}\) and fixed to \(10^{-3}\).
We tuned the sampling temperature from \(\{0.05,0.1,0.15,0.2\}\). The best values were 0.10 for Chengdu and 0.13 for Shanghai.
We balanced the four objectives in Eq.~\eqref{eq:17} by performing a greedy search over a predefined grid for each weight \(w\) from 0.5 to 3. Starting from equal weights, we iteratively adjusted one component at a time based on validation performance (R@1), holding the others fixed, until no further improvement was observed. The final configurations selected from this grid are as follows.
For the \textit{NoGoal} setting, we used \([w_{rep}, w_{rank}, w_{pred}, w_d] = [1.0, 1.0, 1.0, 2.4]\) on the Chengdu dataset, and \([1.3, 2.8, 0.5, 2.9]\) on the Shanghai dataset. In the \textit{GoalD} setting, the selected weights were \([1.0, 1.0, 1.0]\) for Chengdu and \([1.4, 2.1, 1.7]\) for Shanghai. For the \textit{Goal} setting, we used \([2.4, 2.2, 2.8]\) and \([1.9, 1.3, 2.4]\) for Chengdu and Shanghai, respectively. All selected values lie within the predefined grid.

\subsection{Numerical Results of Ablation Analysis}
\label{apx:abl}

Table~\ref{tab:apx_abl} shows the detailed ablation analysis results under the \textit{GoalD} setting.

\renewcommand{\arraystretch}{1.1}
\begin{table}[h]
\centering
\caption{Full ablation analysis results under GoalD.}
\resizebox{\linewidth}{!}{
\begin{tabular}{lccccccccc}
\toprule
\multirow{2}{*}{Chengdu}  & \multicolumn{3}{c}{Link-level} & \multicolumn{6}{c}{Route-level}               \\
                          & R@1      & R@5      & R@10     & R@1   & R@5   & R@10  & M@1   & M@5   & M@10  \\ \midrule
RouteKG                   & 0.916    & 0.979    & 0.988    & 0.815 & 0.953 & 0.974 & 0.815 & 0.866 & 0.869 \\
w/o rerank                & 0.874    & 0.934    & 0.960    & 0.732 & 0.910 & 0.943 & 0.732 & 0.804 & 0.808 \\
w/o relation              & 0.910    & 0.979    & 0.989    & 0.805 & 0.953 & 0.973 & 0.805 & 0.861 & 0.864 \\ \midrule
\multirow{2}{*}{Shanghai} & \multicolumn{3}{c}{Link-level} & \multicolumn{6}{c}{Route-level}               \\
                          & R@1      & R@5      & R@10     & R@1   & R@5   & R@10  & M@1   & M@5   & M@10  \\ \midrule
RouteKG                   & 0.843    & 0.946    & 0.963    & 0.723 & 0.894 & 0.918 & 0.723 & 0.780 & 0.784 \\
w/o rerank                & 0.793    & 0.855    & 0.895    & 0.658 & 0.804 & 0.856 & 0.658 & 0.713 & 0.719 \\
w/o relation              & 0.837    & 0.942    & 0.962    & 0.717 & 0.891 & 0.915 & 0.717 & 0.776 & 0.779 \\ \bottomrule
\end{tabular}
}
\label{tab:apx_abl}
\end{table}

\bibliographystyle{IEEEtran}
\bibliography{main}
\end{document}